\let\ifextended\iftrue
\makeatletter\@ifundefined{ifextended}{\let\ifextended\iffalse}{}\makeatother
\ifextended\documentclass[hidelinks]{siamart251216}\else\documentclass[review]{siamart251216}\fi
\usepackage{amsmath,amssymb,amsfonts}
\usepackage{graphicx}
\usepackage{float}
\usepackage{tikz}
\usetikzlibrary{decorations.pathreplacing}
\usepackage{mathtools}

\newsiamthm{condition}{Condition}
\newsiamremark{remark}{Remark}

\newcommand{\R}{\mathbb{R}}
\newcommand{\Rnn}{\mathbb{R}_{\ge 0}}

\newcommand{\norm}[1]{\left\| #1 \right\|}

\definecolor{cblue}{rgb}{0,0,0}
\definecolor{cred}{rgb}{0,0,0}

\title{Necessary and Sufficient Structural Conditions for the Validity of Lyapunov-Type Exponential Stability Criteria\thanks{\ifextended Extended version of a manuscript submitted to the \emph{SIAM Journal on Control and Optimization}; the main text coincides with the submitted version up to six cross-references to the appendices, which contain the material omitted there for length.\else Submitted to the editors on October 27, 2026.\fi}}
\author{Yixiang Guo\thanks{School of Automation Science and Electrical Engineering, Beihang University, Beijing 100191, People's Republic of China (\email{guoyixiang1234@buaa.edu.cn}, \email{tielin@buaa.edu.cn}).}
\and Lin Tie\footnotemark[2]
\and Quan Quan\thanks{Corresponding author. School of Automation Science and Electrical Engineering, Beihang University, Beijing 100191, People's Republic of China (\email{qq\_buaa@buaa.edu.cn}).}}
\headers{Structural Conditions for Lyapunov-Type Criteria}{Yixiang Guo, Lin Tie, and Quan Quan}

\begin{document}
\maketitle
\begin{abstract}
Whether the exponential decay of a Lyapunov function along trajectories implies exponential stability of the state---an inference we call \emph{migration}---is long assumed in Lyapunov's second method but never characterized at the criterion level. We formalize stability criteria over trajectory families, distinguish admissibility of a criterion (its structural condition suffices for its conclusion) from tightness (its necessity), and establish the necessary and sufficient structural conditions for validity. In logarithmic coordinates the theory reduces to a single two-sided displacement function of the comparison pair: in discrete time, migration holds if and only if a regularity package (excluding failure by infinitely many steps of finite displacement) and a joint lower bound (excluding failure by a single step of unbounded expansion) hold simultaneously. Continuous time strengthens the latter; the two criteria are separated exactly by the minimum step size. Written without its inverse-function formulation, the Panteley--Lor\'ia criterion coincides with our continuous-time conditions on every nondecreasing pair, and loses soundness beyond: it was already complete on the classical regularity class. An envelope construction shows that regularization costs no discriminative power, and radial unboundedness, a standing assumption since Hahn, is dispensable.
\end{abstract}

\begin{keywords}
comparison functions, exponential stability, Lyapunov methods, necessary and sufficient conditions, stability criteria
\end{keywords}

\begin{MSCcodes}
93D23, 93D30, 93D05, 39A30, 34D20
\end{MSCcodes}

\section{Introduction}
\label{sec:introduction}
Lyapunov's second method infers convergence of the state from the decay of a scalar function along trajectories. For exponential stability, one verifies
\[
V(x(t))\le e^{-\beta t}V(x_0)
\]
along trajectories, and hopes to conclude that the state satisfies
\[
\norm{x(t)}\le C\rho^t\norm{x(0)}.
\]
This inference is called \emph{migration}. We prove that whether it is rigorous depends entirely on the function-value properties of the bounding pair: if \(V\) is sandwiched by comparison functions \(\alpha_1,\alpha_2\), i.e., \(\alpha_1(\norm{x})\le V(x)\le\alpha_2(\norm{x})\), then migration holds or fails according to the pair alone. We study the inference itself: given only the exponential decay of \(V\) and no system information, which structural conditions on \(V\), via its bounding pair, make migration hold---and cannot be simplified further (each individually necessary, jointly sufficient)?

This viewpoint requires promoting the criterion from an empirical device to a mathematical object. Section~\ref{sec:prelim} gives the rigorous definition and distinguishes two levels: a criterion is \emph{admissible} if its structural conditions suffice for its conclusion, and \emph{tight} if they are also necessary. Classical criteria are admissible but far from tight; the aim of this paper is a structural condition making the migration criterion both admissible and tight---that is, a necessary and sufficient condition for the criterion's conclusion.

Criteria based on comparison functions go back to Hahn \cite{hahn1967}, following Massera's introduction of comparison functions into stability analysis \cite{massera1956}: if \(V\) admits power bounds of the same order, \(\eta_1\norm{x}^p\le V(x)\le\eta_2\norm{x}^p\), migration holds. He and Wang \cite{he1991} showed that the same-power requirement is not necessary---the only earlier brush with the necessity side, whose answer (``not necessary'') marked boundary hunting as unpromising. (The weak form of the same-order requirement, a bounded ratio \(\alpha_2/\alpha_1\), remains necessary; see Remark~\ref{rem:weak_same_order}.) The Panteley--Lor\'ia criterion \cite{panteley2010} relaxed Hahn's condition to growth and contraction conditions on \(\mathcal{K}_\infty\) pairs, covering variable powers and other non-fixed-order pairs; yet pairs with slow growth or fast decay (effective power tending to zero) remain beyond it---and such pairs exist already within \(\mathcal{K}_\infty\), so the failure is a gap in the conditions' content, not of the class. These criteria share a common structure: sufficient conditions imposed on the pair, leaving a discriminative blind spot---pairs for which existing criteria cannot determine whether migration holds, even though the system may actually be exponentially stable.

Necessity here is distinct from the necessity provided by converse theorems \cite{massera1949,kurzweil1956,lin1996}---existential in form (stability implies the existence of \emph{some} Lyapunov function)---and from the necessity read in \cite{panteley2010} in the same existential sense (via a converse theorem producing a quadratic-bounded Lyapunov function; see Section~\ref{subsec:panteley}). System-level characterizations act on the derivative of \(V\) along the trajectories of a known system \cite{jagt2026}, and \cite{peet2009} shows that some polynomial Lyapunov function always certifies exponential stability on bounded regions. To the best of our knowledge, necessary conditions on a \emph{prescribed} comparison pair---equivalently, necessary conditions for the validity of a migration criterion---have not been studied.

The question has also become practical. Learning-based stability certificates---Lyapunov functions produced by numerical training---come with nothing but numerically estimated bounding pairs \cite{dawson2023}, and whether such a pair certifies anything is precisely a prescribed-pair question; its verification here is dimension-free, since only scalar functions of a scalar variable are involved, unlike sum-of-squares constructions, which scale poorly with the state dimension. Follow-up work on the Panteley--Lor\'ia criterion \cite{panteley2010} has been sparse for fifteen years; the reason is itself instructive: its necessity was read through converse theorems, which produce quadratic-bounded Lyapunov functions on which the growth conditions always hold, so at the existential level the criterion never appears to fail; and its statement through the inverse function \(\alpha_1^{-1}\) imported the \(\mathcal{K}_\infty\) regularity, hiding that the two inequalities are already the exact boundary on the monotone class (Proposition~\ref{prop:pl_equiv}).

The core observation is that the discriminative power of migration can be completely characterized by the function-value properties of the logarithmic comparison functions \(\chi_i(s)=\ln\alpha_i(e^s)\). Taking logarithms transforms multiplicative inequalities into additive ones---in the spirit of the logarithmic twisting of \cite{grune1999}---and all relevant information is then carried by a single two-sided displacement function of the pair. In discrete time, migration holds if and only if this displacement function admits a one-sided linear lower bound at large dips---a regularity package excluding failure by infinitely many steps of finite displacement---together with a joint lower bound at level \(\beta\), which excludes failure by a single step of unbounded expansion. Equivalently, migration holds if and only if the pair satisfies three conditions: bounded closeness, large-scale Lipschitzness, and a joint lower bound; these conditions require no continuity, strict monotonicity, or fixed powers. Continuous time strengthens the joint lower bound, and the two criteria are separated precisely by the minimum step size.

The main contributions are as follows. (i) We prove that the discrete-time strict \(\beta\)-criterion is valid---both admissible and tight---if and only if the comparison pair satisfies the regularity package and the joint lower bound at level \(\beta\) (Theorem~\ref{thm:strict_beta}, Proposition~\ref{prop:packaging}), and we establish the continuous-time counterpart, the two criteria being separated exactly by the minimum step size (Theorems~\ref{thm:general} and~\ref{thm:continuous}); the conditions are the weakest possible within the comparison-pair framework, each individually necessary (Remark~\ref{rem:minimality}). (ii) Written without its inverse-function formulation, the Panteley--Lor\'ia criterion coincides with our continuous-time conditions exactly on the nondecreasing pairs (Proposition~\ref{prop:pl_equiv}), beyond which the direct inequalities cease to be sound (Remark~\ref{rem:pl_sharp}). (iii) An envelope construction polishes every admissible pair into a smooth, nondecreasing, Lipschitz pair at no loss of discriminative power, complemented by an integral-form necessary check (Section~\ref{sec:envelope}); along the way, radial unboundedness---a standing assumption since Hahn---is shown to be dispensable (Section~\ref{subsec:tails}).

The paper is organized as follows. Section~\ref{sec:prelim} formalizes stability criteria over trajectory families and develops the logarithmic characterization on which the paper rests. Section~\ref{sec:conditions} proves the necessary and sufficient conditions for the discrete-time criterion. Section~\ref{sec:continuous} establishes the continuous-time criterion and locates the Panteley--Lor\'ia criterion within it. Section~\ref{sec:envelope} constructs regular envelopes, gives an integral-form necessary check and verifiable classes, and works one example in closed form. Section~\ref{sec:conclusion} concludes. Due to page limitations, several constructions and the numerical examples are omitted\ifextended \ from the journal version; they are collected in the appendices of this extended version\else; they are collected in the extended version of this paper~\cite{guo2026extended}\fi.

\paragraph{Notation} The state space is \(\R^n\); \(r=\norm{x}\) and \(s=\ln r\) are the radial and logarithmic coordinates. Trajectories are denoted by \(\xi\), and \(\Xi\) denotes a trajectory set. The comparison functions are \(\alpha_1,\alpha_2\), with logarithmic counterparts \(\chi_i(s)=\ln\alpha_i(e^s)\); \(g(d,s)=\chi_1(s+d)-\chi_2(s)\) is the displacement function, and \(G(d)=\inf_{s\in\R}g(d,s)\) for \(d\ge0\) is the joint lower envelope. The constants of the conditions are \(R,M\) (regularity package), \(D\) (bounded closeness), and \(\Lambda,r_0\) (large-scale Lipschitzness); \(L,U\) denote envelopes. The letter \(\sigma\) denotes a class-\(\mathcal{KL}\) function in Section~\ref{subsec:comparison} and, locally, the mollification width in Section~\ref{subsec:envelope_construction}; \(\theta\) is the criterion parameter in Section~\ref{subsec:criteria}. Envelope stages carry subscripts \(\mathrm{mon}\) (monotonized), \(\mathrm{sc}\) (semi-continuous), \(\mathrm{lip}\) (Lipschitzized), and \(\sigma\) (mollified).

\section{Stability criteria: rigorous formulation and analysis}
\label{sec:prelim}

The questions of this paper concern criteria, not any particular system: whether the exponential decay of a Lyapunov function along trajectories can be transferred to exponential contraction of the state, and what structure of the bounding functions makes this transfer exact. To ask such questions precisely, the notion of a criterion must itself be made a mathematical object. This section supplies the formulation (Subsection~\ref{subsec:criteria}) and then reduces migration to a property of the comparison pair in logarithmic coordinates (Subsection~\ref{subsec:analysis}). The trajectory classes below contain arbitrary curves, not only solutions of differential equations; the criteria are therefore stronger than the corresponding system-level statements---they apply to every system admitting such a \(V\)---while their necessity is a criterion-level property (Remark~\ref{rem:attainability}).

\subsection{Stability criteria}
\label{subsec:criteria}

Let \(\Xi\) denote a set of trajectories: its elements \(\xi\in\Xi\) are curves---arbitrary functions of the time variable, with no continuity or measurability imposed---and \(\xi(t)\) denotes the state of \(\xi\) at time \(t\). The time domain is left unspecified---a trajectory set may contain discrete-time trajectories (\(t\in\mathbb{N}\)), continuous-time ones (\(t\ge0\)), or a mixture of both; no generating system is assumed, and trajectories are denoted by \(\xi\) precisely because no solution map is available.

When one considers a class of systems or a class of trajectories, a general parameter set \(\Theta\) serves to label the subclasses or individual members: for instance, for uncontrolled linear systems with an equilibrium at the origin, the parameter set is \(\Theta=\{A:A\in\R^{n\times n}\}\). To each parameter corresponds a trajectory set, which leads to the following definition.

\begin{definition}[Family of trajectory sets]\label{def:family}
Let \(\Theta\) be a general parameter set. A \emph{family of trajectory sets} based on \(\Theta\) is
\[
\Omega=\{\Xi_\theta:\theta\in\Theta\},
\]
where each \(\Xi_\theta\) is a trajectory set as above.
\end{definition}

\begin{remark}[What the parameter can mean]\label{rem:parameter}
For a fixed system, \(\Theta\) is a singleton whose member is the set of all trajectories of that system; in general, a parameter may encode a system parameter, a class of systems, or an additional requirement imposed by the criterion. The definitions are deliberately neutral about this. One could additionally impose a soundness requirement---that each \(\Xi_\theta\) cover all trajectories of at least one system from a background class---but we do not: no proof uses it, and the criteria are studied as abstract objects over trajectory families. The feasible sets of the criteria studied below are nonetheless nonempty---curves that jump to the origin are feasible---except at the trivial initial value in the strict case, which is vacuous by the conventions below.
\end{remark}

\begin{definition}[Global exponential stability criterion]\label{def:criterion_param}
A \emph{global exponential stability criterion} consists of a \emph{feasible trajectory condition}---a family \(\Omega=\{\Xi_\theta:\theta\in\Theta\}\) of feasible trajectory sets---and a \emph{conclusion}: for every \(\theta\in\Theta\), there exist constants \(C_\theta\ge1\) and \(\rho_\theta\in(0,1)\) such that every \(\xi\in\Xi_\theta\) satisfies \(\norm{\xi(t)}\le C_\theta\rho_\theta^t\norm{\xi(0)}\).
\end{definition}
The family \(\Omega\) fixes the scope of the conclusion: the larger each \(\Xi_\theta\) and the richer \(\Theta\), the more general the criterion.

A criterion is called \emph{admissible} if its conditions suffice to yield its conclusion, i.e., it can be used in its intended scope, and \emph{tight} if the conditions are moreover necessary. A necessary and sufficient condition for a criterion is, by definition, one for which the criterion is both admissible and tight. A criterion designed via Lyapunov's second method typically rests on an \emph{evolution condition} on \(V\) along trajectories, \emph{structural conditions on \(V\)} itself (power-type bounds, positive definiteness), and \emph{system structural conditions} (linearity, affinity). The criteria studied in this paper use no system structural condition and fix the evolution condition as exponential decay of \(V\) along trajectories; all discriminative power is thereby reduced to the structural conditions on \(V\). Note that a criterion is defined relative to a trajectory family: its failure does not by itself imply that a particular system fails to be exponentially stable (for a specific class of systems, an exactly equivalent criterion may exist, at the price of generality). The system-free formulation is deliberate: it isolates what the comparison pair alone discriminates, which is precisely the data that numerically trained Lyapunov certificates provide. We now give the rigorous definition of the criterion studied here; part of its feasible trajectory condition is fixed and written into the conclusion.

\begin{definition}[Migration criterion]\label{def:criterion}
The \emph{migration criterion} is the global exponential stability criterion based on exponential decay of a Lyapunov function: its structural condition restricts \(V\) itself, requiring no system information, and its conclusion runs over \(\Theta=(0,+\infty)\) with \(\Xi_\theta=\{\xi:V(\xi(t))\le e^{-\theta t}V(\xi(0))\}\): for every \(\theta>0\), there exist constants \(C_\theta\ge1\) and \(\rho_\theta\in(0,1)\) such that every \(\xi\in\Xi_\theta\) satisfies \(\norm{\xi(t)}\le C_\theta\rho_\theta^t\norm{\xi(0)}\).
\end{definition}
In the discrete-time case, \(t\in\mathbb{N}\); in the continuous-time case, \(t\ge0\). This definition and its derivatives will simply be called \emph{criteria}, with the time setting prefixed to the name (e.g., the discrete criterion).

Although \(V\) is called a Lyapunov function, it is not required to meet all the conditions of the classical Lyapunov criterion: positive definiteness, together with the sphere bounds \eqref{eq:annulus}, suffices for the characterization (and continuity enters no proof); in particular, no radial unboundedness is imposed (see the discussion of tail requirements at the end of Subsection~\ref{subsec:analysis}). The problem of this paper can now be stated: which structural conditions are both admissible and tight?

\subsection{Equivalent characterization of the criterion}
\label{subsec:analysis}

\paragraph{Continuous versus discrete time}
The conclusion of the criterion may be required in discrete time or in continuous time. Since the decay rate is arbitrary, one may expect the discrete-time criterion to be equivalent to a continuous-time criterion with a truncated initial segment; this is indeed the case, and a weakened criterion serves as the bridge.

\begin{definition}[Weakened continuous-time criterion]\label{def:criterion_weak}
The two-part criterion of Definition~\ref{def:criterion}, with the conclusion required only beyond an initial segment: \(\Theta=\{(\beta,\varepsilon):\beta>0,\ \varepsilon>0\}\), \(\Xi_\theta=\{\xi:V(\xi(t))\le e^{-\beta t}V(\xi(0))\}\), and for every \(\theta=(\beta,\varepsilon)\) there exist \(C_\theta\ge1\) and \(\rho_\theta\in(0,1)\) such that every \(\xi\in\Xi_\theta\) satisfies \(\norm{\xi(t)}\le C_\theta\rho_\theta^t\norm{\xi(0)}\) for \(t\ge\varepsilon\).
\end{definition}
This criterion is equivalent to the discrete criterion (Theorem~\ref{thm:cont_disc}): the discrete-time criterion is the continuous-time one minus the initial segment, and the missing segment is restored in Section~\ref{sec:conditions} by a separate analysis.

The general criterion decomposes into a family of criteria, one for each decay rate; we first define the constituent.

\begin{definition}[\(\beta\)-criterion and strict \(\beta\)-criterion]\label{def:criterion_strick}
The two-part criterion of Definition~\ref{def:criterion} with a fixed decay rate: for fixed \(\beta>0\), \(\Xi=\{\xi:V(\xi(t))\le e^{-\beta t}V(\xi(0))\}\) (for the strict \(\beta\)-criterion, with the strict inequality \(V(\xi(t))<e^{-\beta t}V(\xi(0))\)), and the conclusion is \(\norm{\xi(t)}\le C\rho^t\norm{\xi(0)}\) for \(t>0\) with constants \(C\ge1\) and \(\rho\in(0,1)\).
\end{definition}
The case \(t=0\) is trivial throughout the paper, so we henceforth assume \(t>0\). To distinguish the criteria introduced so far, the criterion of Definition~\ref{def:criterion} is henceforth called the \emph{general criterion}.

\paragraph{Composition of the criterion}
We first record an evident transfer: if a structural condition makes the strict \(\beta\)-criterion admissible, it also makes the \(2\beta\)-criterion admissible, since \(\Xi_{2\beta}\subseteq\Xi_\beta^{\mathrm{s}}\) away from the trivial trajectory (for \(V(\xi(0))>0\), \(V(\xi(k))\le e^{-2\beta k}V(\xi(0))<e^{-\beta k}V(\xi(0))\)); at \(V(\xi(0))=0\) the zero trajectory is not strictly feasible (\(0<0\) fails), but it satisfies every homogeneous exponential bound trivially. Since \(\beta\) is arbitrary and the general criterion is usable exactly when the \(\beta\)-criterion is usable for every \(\beta>0\), the general criterion is admissible if and only if the strict \(\beta\)-criterion is admissible for every \(\beta>0\); tightness (necessity) transfers along the same inclusion.

Thus the strict \(\beta\)-criterion is the constituent of the general criterion. Moreover, only the worst feasible trajectories constrain the conclusion, which leads to an equivalent form.

\paragraph{An equivalent form of the conclusion}
Grouping trajectories by the initial norm \(r=\norm{x_0}\), the conclusion reads equivalently: there exist constants \(C_\beta\ge1\) and \(\rho_\beta\in(0,1)\) such that, for every \(r\ge0\),
\begin{equation}\label{eq:worst}
\sup_{\norm{x_0}=r}\ \sup_{\xi:\,\xi(0)=x_0}\norm{\xi(t)}\ \le\ C_\beta\rho_\beta^t\,r,\qquad t>0.
\end{equation}
Both sides of this form involve only the norm of the state, i.e., the problem is radially symmetric; this suggests a radial symmetrization of \(V\).

\paragraph{Radial bounds and the characterization}
Given a Lyapunov function \(V:\R^n\to\Rnn\), define its radially symmetric upper and lower bounds
\begin{equation}\label{eq:radial}
\alpha_1(r)\coloneqq\inf_{\norm{x}=r}V(x),\qquad
\alpha_2(r)\coloneqq\sup_{\norm{x}=r}V(x),\quad r\ge0.
\end{equation}
By definition, the sandwich condition \(\alpha_1(\norm{x})\le V(x)\le\alpha_2(\norm{x})\) follows directly. By the basic requirement that \(V\) certify asymptotic stability, \(\alpha_1(0)=\alpha_2(0)=0\); no radial unboundedness is imposed on \(V\). Positive definiteness alone does not prevent a discontinuous \(V\) from attaining arbitrarily small or arbitrarily large values on a sphere, so we assume explicitly
\begin{equation}\label{eq:annulus}
0<\alpha_1(r)\le\alpha_2(r)<+\infty,\qquad r>0,
\end{equation}
i.e., \(V\) has a positive lower bound and a finite upper bound on every sphere; the logarithmic pair \eqref{eq:chi} below is then finite-valued. (The stronger classical convention, boundedness on compact annuli, is not needed: no proof below uses it.)

We adopt the convention that the supremum of an empty set is read as \(0\) for norm sets (which consist of nonnegative numbers) and as \(-\infty\) for logarithmic level sets, so that the characterizations below hold vacuously when the relevant set is empty; in particular, \(\Gamma_\beta=-\infty\) below means that no state is admissible and the bound is vacuous (accordingly, \(e^{-\infty}=0\) in the norm coordinates).

\begin{theorem}[Equivalent characterization of the strict \(\beta\)-criterion conclusion]\label{thm:comp_char}
The conclusion \eqref{eq:worst} of the strict \(\beta\)-criterion is equivalent to
\begin{equation}\label{eq:comp_char}
\sup\{\norm{x}:\alpha_1(\norm{x})<e^{-\beta t}\alpha_2(r)\}\le C_\beta\rho_\beta^t\,r,\quad t>0,\ r\ge0.
\end{equation}
\end{theorem}

\begin{proof}
Fix \(r\ge0\) and an initial state \(x_0\) with \(\norm{x_0}=r\). If \(V(x_0)=0\), then \(x_0=0\) and the strict decay reads \(V(\xi(t))<0\), so no trajectory from \(x_0\) is feasible; the set \(\{x:V(x)<e^{-\beta t}V(x_0)\}\) is likewise empty, so both suprema vanish under the conventions stated above. Assume henceforth that \(V(x_0)>0\). The inner supremum in \eqref{eq:worst} is taken over all trajectories from \(x_0\), so
\[
\sup_{\xi:\,\xi(0)=x_0}\norm{\xi(t)}=\sup\{\norm{x}:V(x)<e^{-\beta t}V(x_0)\}.
\]
Indeed, suppose that some state \(x\) with \(V(x)<e^{-\beta t}V(x_0)\) were attained at time \(t\) by no trajectory from \(x_0\). The curve \(\xi\) with \(\xi(\tau)=x\) for \(0<\tau\le t\) and \(\xi(\tau)=0\) for \(\tau>t\) satisfies the decay constraint---since \(V(x)<e^{-\beta t}V(x_0)\le e^{-\beta\tau}V(x_0)\) for \(\tau\le t\), and \(V(0)=0\) thereafter---and attains \(x\) at time \(t\), a contradiction. In discrete time the construction reads \(\xi(0)=x_0\), \(\xi(k)=x\) for \(1\le k\le t\), and \(\xi(k)=0\) for \(k>t\), and the decay check \(V(x)<e^{-\beta t}V(x_0)\le e^{-\beta k}V(x_0)\) for \(k\le t\) applies verbatim. Passing to norm sets, \(\{\norm{x}:V(x)<e^{-\beta t}V(x_0)\}=\{\norm{x}:\alpha_1(\norm{x})<e^{-\beta t}V(x_0)\}\). Taking the supremum over \(\norm{x_0}=r\) then replaces \(V(x_0)\) by \(\alpha_2(r)\):
\[
\sup_{\norm{x_0}=r}\sup\{\norm{x}:\alpha_1(\norm{x})<e^{-\beta t}V(x_0)\}
=\sup\{\norm{x}:\alpha_1(\norm{x})<e^{-\beta t}\alpha_2(r)\}.
\]
Indeed, the direction ``\(\le\)'' follows from \(V(x_0)\le\alpha_2(r)\); for ``\(\ge\)'', denote the right-hand side by \(S\), and for any \(\delta>0\) pick a radius \(\bar r\) exceeding \(S-\delta\) with \(\alpha_1(\bar r)<e^{-\beta t}\alpha_2(r)\); then, since the strict inequality leaves a gap, pick \(\norm{x_0}=r\) with \(V(x_0)\) so close to the supremum \(\alpha_2(r)\) that \(\alpha_1(\bar r)<e^{-\beta t}V(x_0)\) as well; letting \(\delta\to0\) yields the claim.
\end{proof}

\begin{remark}[Role of the system-free setting]\label{rem:attainability}
The equivalence of Theorem~\ref{thm:comp_char} rests on the absence of system structure: every state compatible with the decay constraint is attained by some feasible trajectory (the contradiction construction in the proof). For a specific system, the level-set supremum overestimates the worst-case state, and the necessity direction collapses; necessity is therefore a criterion-level property, not a system-level one.
\end{remark}

\begin{remark}[Comparison functions]\label{rem:comparison}\label{subsec:comparison}
The pair \((\alpha_1,\alpha_2)\) in \eqref{eq:radial} is a comparison-function pair in the classical sense, although we do not require the classical regularity. Recall that \(\alpha:\Rnn\to\Rnn\) is of class \(\mathcal{K}\) if it is continuous, strictly increasing, and \(\alpha(0)=0\), and of class \(\mathcal{K}_\infty\) if in addition \(\alpha(s)\to+\infty\) as \(s\to+\infty\); \(\sigma:\Rnn\times\Rnn\to\Rnn\) is of class \(\mathcal{KL}\) if \(\sigma(\cdot,t)\in\mathcal{K}\) and \(\sigma(s,\cdot)\) decreases to \(0\). The classical comparison function method sandwiches \(V\) between \(\alpha_1,\alpha_2\in\mathcal{K}_\infty\) and transfers the decay \(V(\xi(t))\le\sigma(V(x_0),t)\), \(\sigma\in\mathcal{KL}\), to the state via \(\norm{\xi(t)}\le\alpha_1^{-1}\bigl(\sigma(\alpha_2(\norm{x_0}),t)\bigr)\). We impose almost no regularity on the pair itself; this covers variable powers, slow growth, and fast decay beyond the classical framework; the price is that \(\alpha_1^{-1}\) may fail to exist, so the transfer chain breaks and is replaced by the level-set supremum below. Related discrete-time comparison-function methods can be found in \cite{jiang2001,teel2010,pepe2025,lakshmikantham1990}; a systematic summary of comparison functions is \cite{kellett2014}.
\end{remark}

\begin{remark}[Realizability of pairs]\label{rem:realizability}
The pair conventions are not vacuous: every logarithmic comparison pair arises from an explicit function, and each regularity property of the pair transfers. For a given pair, write \(\alpha_i(r)\coloneqq\exp(\chi_i(\ln r))\) for \(r>0\) and set
\[
V(x)\coloneqq\alpha_1(\norm{x})+\bigl(\alpha_2(\norm{x})-\alpha_1(\norm{x})\bigr)\cos^2\theta,\qquad x\neq0,
\]
with \(\theta\) the polar angle of \(x\), and \(V(0)\coloneqq0\) under the left tail below. Since \(\cos^2\theta\) attains every value of \([0,1]\) on each circle, the radial bounds \eqref{eq:radial} of this \(V\) are exactly \((\alpha_1,\alpha_2)\): no hypothesis on the pair is needed, and \(V\) is positive definite off the origin automatically, finite-valued \(\chi_i\) meaning \(\alpha_i>0\) for \(r>0\). The correspondence is a dictionary: \(V\) is continuous exactly when the \(\chi_i\) are; \(V\) is smooth on \(\R^2\setminus\{0\}\) exactly when the \(\chi_i\) are smooth, \(s=\ln r\) being a smooth bijection; continuity of \(V\) at the origin with \(V(0)=0\) is exactly the left tail \(\chi_2(-\infty)=-\infty\); and radial unboundedness is exactly \(\chi_1(+\infty)=+\infty\) (Section~\ref{subsec:tails}). Dynamics realizing exact decay are likewise explicit: in discrete time, for a continuous pair with proper tails, the radial map \(F(x)\coloneqq\rho(\norm{x})\,x/\norm{x}\), where \(\rho(r)\coloneqq\sup\{\varrho\ge0:\alpha_2(\varrho)\le e^{-\beta}\alpha_1(r)\}\), satisfies \(V(F(x))\le e^{-\beta}V(x)\); in continuous time, if the pair is moreover strictly increasing, the level-set parametrization \(\xi(t)\coloneqq w\bigl(e^{-\beta t}V(x_0),\theta_0\bigr)(\cos\theta_0,\sin\theta_0)\), where \(w(\cdot,\theta_0)\) inverts the strictly increasing function \(r\mapsto V(r\cos\theta_0,r\sin\theta_0)\), defines a continuous semiflow along which \(V\) decays exactly as \(e^{-\beta t}V(x_0)\), consistent with the trajectory-level framework of this paper. Pairs failing the tail requirements do not arise from any continuous Lyapunov function; those failing only strict monotonicity---the plateau pair of \ifextended Appendix~\ref{app:strict}\else the extended version~\cite{guo2026extended}\fi, for instance---do arise as envelopes of explicit Lyapunov functions, but admit dynamics only through the trajectory-level admissibility of this paper (jump curves), not through a continuous semiflow with exact decay.
\end{remark}

We now pass to logarithmic coordinates, which turn multiplicative inequalities into additive ones and thereby reduce the complexity of the characterization.

\paragraph{Logarithmic coordinates}
Define the logarithmic comparison functions
\begin{equation}\label{eq:chi}
\chi_i(s)\coloneqq\ln\alpha_i(e^s),\quad i=1,2,\quad s\in\R.
\end{equation}
Throughout the paper, \(\chi_1,\chi_2:\R\to\R\) are finite-valued and satisfy \(\chi_1\le\chi_2\) (both hold automatically for pairs arising from \eqref{eq:radial}); such a pair is called a \emph{logarithmic comparison pair}. Define the level-set supremum function
\begin{equation}\label{eq:Gamma}
\Gamma_\beta(s_0,t,\chi_1,\chi_2)\coloneqq\sup\{s\in\R:\chi_1(s)<\chi_2(s_0)-\beta t\},
\end{equation}
where \(s_0\coloneqq\ln\norm{x_0}\) denotes the starting point in logarithmic coordinates; we abbreviate \(\Gamma_\beta(s_0,t)\). Under these empty-set conventions, \(\Gamma_\beta\) is nonincreasing in \(t\), and
\begin{equation}\label{eq:Gamma_bound}
\ln\norm{\xi(t)}\le\Gamma_\beta(s_0,t),\quad t>0.
\end{equation}
Level-set constructions have appeared in converse Lyapunov theorems \cite{massera1949,kurzweil1956,yoshizawa1966}; related tools include smoothing techniques \cite{wilson1969,lin1996}, discrete-time smooth Lyapunov functions \cite{kellett2004}, converse theorems for difference equations \cite{gordon1972}, and convergence characterizations for discrete-time systems \cite{tran2019}. The recent work \cite{jagt2026} gives a system-level characterization. The \(\Gamma_\beta\) here does not depend on particular trajectories; it is determined solely by the pair \((\alpha_1,\alpha_2)\) and reduces the discriminative power of the criterion to function-value properties of the comparison functions.

All information carried by the pair is condensed into a single two-sided displacement function
\begin{equation}\label{eq:g}
g(d,s)\coloneqq\chi_1(s+d)-\chi_2(s),\qquad d,s\in\R,
\end{equation}
where \(d\) is \emph{not} restricted to be nonnegative (negative \(d\) controls the rises of \(\chi_1\) in the necessity proof), together with its one-sided summary, the \emph{joint lower envelope}
\begin{equation}\label{eq:G}
G(d)\coloneqq\inf_{s\in\R}g(d,s),\qquad d\ge0,
\end{equation}
with \(G_{\min}\coloneqq\inf_{d\ge0}G(d)\) and \(G_{\lim}\coloneqq\liminf_{d\to+\infty}G(d)\). The two objects have complementary roles: the proofs require the two-sided \(g\), whereas the second condition of each criterion involves only \(G\). The level-set supremum reads
\begin{equation}\label{eq:Gamma_g}
\Gamma_\beta(s_0,t)=s_0+\sup\{d\in\R:g(d,s_0)<-\beta t\},
\end{equation}
so the criterion's conclusion will take the form: a drop of \(\beta t\) can be realized only by a displacement of at most \(T_\beta-t\eta_\beta\).

\begin{definition}[Logarithmic strict \(\beta\)-criterion]\label{def:criterion_logarithm}
The structural condition restricts the pair \((\chi_1,\chi_2)\) itself, requiring no system information; the conclusion requires constants \(T_\beta\ge0\) and \(\eta_\beta>0\) such that for every \(s_0\in\R\),
\begin{equation}\label{eq:propertyP}
\Gamma_\beta(s_0,t)\le s_0+T_\beta-t\eta_\beta,
\end{equation}
equivalently, in terms of the displacement function \eqref{eq:g},
\begin{equation}\label{eq:propertyP_g}
g(d,s_0)<-\beta t\ \Longrightarrow\ d\le T_\beta-t\eta_\beta.
\end{equation}
In the discrete-time case, \(t\in\mathbb{N}^{*}\); in the continuous-time case, \(t>0\).
\end{definition}

The conclusion \eqref{eq:propertyP} is the logarithmic form of \eqref{eq:comp_char}. Incorporating the decay rate into the conclusion via the level-set supremum makes the criterion's conditions independent of the time setting.

\begin{theorem}[Continuous-discrete equivalence]\label{thm:cont_disc}
The weakened continuous-time general criterion is equivalent to the discrete-time general criterion, i.e., the two criteria share the same structural conditions.
\end{theorem}

\begin{proof}
\emph{Continuous \(\Rightarrow\) discrete}: take \(\varepsilon=1\). Every discretely feasible trajectory \(\eta\) extends to a continuously feasible one by \(\xi(k)\coloneqq\eta(k)\) at integers and \(\xi(t)\coloneqq0\) otherwise: at noninteger times \(V(\xi(t))=0\le e^{-\beta t}V(\eta(0))\), and at integers the decay holds by the discrete feasibility of \(\eta\). The conclusion at \(t\ge1\), applied to \(\xi\) and restricted to integers, is the discrete-time conclusion.

\emph{Discrete \(\Rightarrow\) continuous}: fix \(\varepsilon>0\). For \(t\ge\varepsilon\), write \(t=(k+\delta)\varepsilon\) with \(k=\lfloor t/\varepsilon\rfloor\ge1\) and \(\delta\in[0,1)\). By the monotonicity of \(\Gamma_\beta\) in \(t\), and the discrete-time conclusion applied to the sampled trajectory at step \(k\) with per-step rate \(\beta\varepsilon\),
\[
\Gamma_\beta(s_0,t)\le\Gamma_\beta(s_0,k\varepsilon)
\le s_0+T_{\beta\varepsilon}-k\eta_{\beta\varepsilon}
\le s_0+T_{\beta\varepsilon}-\frac{t}{2\varepsilon}\eta_{\beta\varepsilon},
\]
since \(k\ge(k+\delta)/2=t/(2\varepsilon)\). Taking \(T_{\beta,\varepsilon}=T_{\beta\varepsilon}\) and \(\eta_{\beta,\varepsilon}=\eta_{\beta\varepsilon}/(2\varepsilon)\) yields the weakened continuous-time criterion.
\end{proof}

\paragraph{On the classical tail requirements}
\label{subsec:tails}
Classical formulations impose, beyond \(\chi_1\le\chi_2\), the tail requirements \(\chi_2(-\infty)=-\infty\) (the analogue of \(V(0)=0\)) and \(\chi_1(+\infty)=+\infty\) (radial unboundedness), as well as boundedness on compact intervals. None of them enters the characterization of Section~\ref{sec:conditions}, whose proofs use nothing beyond the pair definition. Their actual roles are as follows. (i) The left tail corresponds to the continuity of \(V\) at the origin: it makes the level sets in \eqref{eq:Gamma} nonempty for arbitrarily deep thresholds, and it is what allows the pair to arise from a (continuous) Lyapunov function. Dropping it is nonetheless legitimate, and the characterizations above are unaffected: \(\chi\) bounded below at \(-\infty\) means that \(V\) admits an essential infimum \(c>0\) on a punctured neighborhood of the origin, i.e., \(V\) is discontinuous there. Since \(V(0)=0\), the origin always satisfies the strict decay, and once the threshold falls below \(c\) the only admissible state nearby is the origin itself: the criterion then certifies exact finite-time arrival at the origin, the characterizations themselves unchanged. The two degeneracies compose: dropping the right tail (ii) gives uniform settling into every neighborhood of the origin, and dropping the left tail gives strict arrival, so a pair with neither tail certifies strict fixed-time stability. The envelope construction of Section~\ref{sec:envelope} assumes the left tail only for the correspondence with a continuous Lyapunov function; the finiteness of the monotone envelopes already follows from the uniform lower bound. (ii) The right tail is dispensable altogether: with \(\chi_1=\chi_2=h\), where \(h(s)=s\) for \(s\le0\) and \(h(s)=1-e^{-s}\) for \(s>0\), \eqref{eq:propertyP} holds with \(T=0\) and \(\eta=\beta\) (since \(h^{-1}\) has slope at least \(1\)), although \(\chi_1(+\infty)=1\). Bounded envelopes even enjoy a uniform settling time into every neighborhood of the origin, \(T_\varepsilon\le\beta^{-1}\ln\bigl(\sup V/\alpha_1(\varepsilon)\bigr)\), a property the classical framework excludes from the outset by requiring radial unboundedness. A caveat: in continuous time, a \(V\) spatially constant on an open region is incompatible with \(\dot V\le-\beta V\), so this relaxation is most natural in discrete time. (iii) Compact-interval boundedness is automatic under the regularity package: a finite-valued function obeying the large-scale Lipschitz condition is bounded on compact intervals.

\section{The discrete-time criterion}
\label{sec:conditions}

By Section~\ref{sec:prelim}, the migration criterion reduces to properties of the displacement function \(g\) of \eqref{eq:g}, with the strict \(\beta\)-criterion as the constituent to study first. This section states the structural conditions, together with their failure mechanisms and an equivalent working form on the pair, and proves that they make the discrete-time criteria both admissible and tight---that is, necessary and sufficient---first for the strict \(\beta\)-criterion, where a single contradiction argument settles displacements of both signs, and then for the general criterion.

\subsection{The discrete-time strict \(\beta\)-criterion: necessary and sufficient conditions}
\label{subsec:necessity}\label{sec:discrete}

\begin{condition}[Regularity package]\label{cond:package}
There exist \(R>0\) and \(M>0\) such that, for every \(s\in\R\) and every \(d\in\R\),
\begin{equation}\label{eq:U1}
g(d,s)\le-R\ \Longrightarrow\ g(d,s)\ge-M|d|.
\end{equation}
\end{condition}

\begin{condition}[Joint lower bound at level \(\beta\)]\label{cond:joint_lower}
With the joint lower envelope \(G\) of \eqref{eq:G}, require that there exist \(d_\beta>0\) such that
\[
G(d)\ge-\beta,\qquad d\ge d_\beta.
\]
\end{condition}

\begin{figure}[tb]
\centering
\begin{tikzpicture}[scale=0.48,yscale=0.82]
\begin{scope}
    \draw[->,very thick] (-0.8,0) -- (6.4,0) node[right] {\(s\)};
    \draw[->,very thick] (0,-2.0) -- (0,3.65) node[above] {\(\chi\)};
    \draw[very thick,cblue,rounded corners=3pt] (-0.5,0.05) -- (0.4,0.7) -- (1.5,1.5) -- (4.6,1.5) -- (5.7,2.6);
    \node[right] at (5.7,2.6) {\(\chi_1\)};
    \draw[very thick,black,dashed,rounded corners=3pt] (-0.5,0.15) -- (0.4,1.05) -- (1.5,2.7) -- (4.6,2.7) -- (5.7,3.55);
    \node[right] at (5.7,3.55) {\(\chi_2\)};
    \draw[dashed,gray,line width=1pt] (1.5,-0.7) -- (1.5,1.5);
    \draw[dashed,gray,line width=1pt] (4.6,-0.7) -- (4.6,1.5);
    \node[anchor=east] at (1.35,-0.4) {\(a\)};
    \node[anchor=west] at (4.75,-0.4) {\(b\)};
    \draw[<->,very thick] (2.1,1.5) -- (2.1,2.7) node[above]{\small \(h\)};
    \draw[<->,very thick] (1.5,-0.7) -- (4.6,-0.7) node[midway,below]{\small \(b-a\)};
    \node at (2.8,-2.55) {\small (a)};
\end{scope}
\draw[->,line width=1.4pt] (7.5,0.55) -- (9.9,0.55);
\begin{scope}[shift={(11.8,0)}]
    \draw[->,very thick] (-0.8,0) -- (6.4,0) node[right] {\(d\)};
    \draw[->,very thick] (0,-2.0) -- (0,3.65);
    \node[above] at (0,3.65) {\(g(\cdot,a)\)};
    \draw[dashed,black,line width=1pt] (0.2,-0.55) -- (6.1,-0.55);
    \draw[very thick] (-0.07,-0.55) -- (0.07,-0.55);
    \node[left] at (-0.09,-0.55) {\small \(-\beta\)};
    \draw[very thick,cblue,rounded corners=4pt] (-0.6,0.55) -- (-0.2,0.7) -- (0.15,0.75) -- (0.55,0.65) -- (0.8,-1.05) -- (5.05,-1.05) -- (5.5,0.75) -- (6.1,1.0);
    \draw[<->,very thick] (2.9,-1.05) -- (2.9,0) node[above]{\small \(h\)};
    \draw[<->,very thick] (0.8,-1.35) -- (5.05,-1.35) node[right]{\small \(b-a\)};
    \node at (2.8,-2.55) {\small (b)};
\end{scope}
\end{tikzpicture}
\caption{Failure of Condition~\ref{cond:joint_lower}: a single step of unbounded expansion. (a) Geometric source in the pair: a plateau of \(\chi_1\) on \([a,b]\), with \(\chi_2\) higher than the plateau start by \(h\); the state crosses the plateau in one step while \(V\) decreases only by the fixed gap \(h\). (b) On the displacement function, \(g(d,a)=-h\) throughout the plateau, so unbounded plateau lengths force \(G(d)\le-h<-\beta\) at arbitrarily large \(d\), violating the threshold requirement.}
\label{fig:U2_fail}
\end{figure}
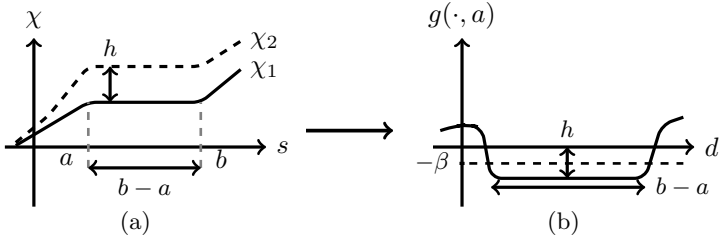

The two conditions correspond to the two ways migration can fail: Condition~\ref{cond:package} excludes failure by \emph{infinitely many steps of finite displacement} (a drop deeper than \(R\) requires a displacement of proportional size), and Condition~\ref{cond:joint_lower} excludes failure by \emph{a single step of unbounded expansion} (the plateau mechanism). The two are dual on the \(g\)-landscape---a well of unbounded depth over bounded width versus a trench of bounded depth over unbounded width; see Figures~\ref{fig:U1_fail} and \ref{fig:U2_fail}.

\begin{figure}[tb]
\centering
\begin{tikzpicture}[scale=0.48,yscale=0.82]
\begin{scope}
    \draw[->,very thick] (-0.8,0) -- (6.4,0) node[right] {\(s\)};
    \draw[->,very thick] (0,-2.0) -- (0,3.65) node[above] {\(\chi\)};
    \draw[very thick,black,dashed] (-0.5,0.62) -- (0.4,1.05) -- (5.7,3.55);
    \node[right] at (5.7,3.55) {\(\chi_2\)};
    \draw[very thick,cblue,rounded corners=2pt] (-0.5,0.15) -- (0.4,0.6) -- (3.2,2.0) -- (3.6,0.5) -- (4.0,2.05) -- (5.7,2.7);
    \node[right] at (5.7,2.7) {\(\chi_1\)};
    \draw[dashed,gray,line width=1pt] (3.2,-0.5) -- (3.2,2.0);
    \draw[dashed,gray,line width=1pt] (4.0,-0.5) -- (4.0,2.05);
    \draw[<->,very thick] (3.2,-0.5) -- (4.0,-0.5) node[midway,below]{\small \(p_n\)};
    \draw[line width=1pt,gray] (4.0,0.5) -- (4.35,0.5);
    \draw[line width=1pt,gray] (4.0,2.05) -- (4.35,2.05);
    \draw[<->,very thick] (4.45,2.05) -- (4.45,0.5) node[midway,right]{\small \(H_n\)};
    \node at (2.8,-2.55) {\small (a)};
\end{scope}
\draw[->,line width=1.4pt] (7.5,0.55) -- (9.9,0.55);
\begin{scope}[shift={(11.8,0)}]
    \draw[->,very thick] (-0.8,0) -- (6.4,0) node[right] {\(d\)};
    \draw[->,very thick] (0,-2.0) -- (0,3.65);
    \node[above] at (0,3.65) {\(g(\cdot,s_n)\)};
    \draw[dashed,gray,line width=1pt] (0.15,-0.05) -- (5.6,-1.9);
    \draw[dashed,gray,line width=1pt] (0.15,-0.09) -- (3.4,-2.1);
    \node[gray,below right] at (5.35,-1.85) {\small \(-M|d|\)};
    \draw[dashed,black,line width=1pt] (0.3,-0.95) -- (5.9,-0.95);
    \draw[very thick] (-0.07,-0.95) -- (0.07,-0.95);
    \node[left] at (-0.09,-0.95) {\small \(-R\)};
    \draw[very thick,cblue,rounded corners=2pt] (-0.6,0.1) -- (-0.2,0.35) -- (0.2,0.9) -- (1.2,1.1) -- (1.33,-1.5) -- (1.46,1.15) -- (2.6,1.45) -- (2.73,-1.72) -- (2.86,1.5) -- (6.0,2.05);
    \draw[->,very thick,cblue] (1.33,-1.5) -- (1.33,-1.85);
    \draw[->,very thick,cblue] (2.73,-1.72) -- (2.73,-1.95);
    \node at (2.8,-2.55) {\small (b)};
\end{scope}
\end{tikzpicture}
\caption{Failure of Condition~\ref{cond:package}: infinitely many steps of finite displacement. (a) Geometric source in the pair: \(\chi_1\) develops dips of unbounded depth \(H_n\) over spans \(p_n\) with \(H_n/p_n\to+\infty\) (\(p_n=0\) being an unbounded gap \(\chi_2-\chi_1\)). (b) On the displacement function, the wells of \(g(\cdot,s_n)\) dive below \(-R\) with bounded width, eventually crossing every line \(-M|d|\); no constants \(R,M\) exist for \eqref{eq:U1}.}
\label{fig:U1_fail}
\end{figure}
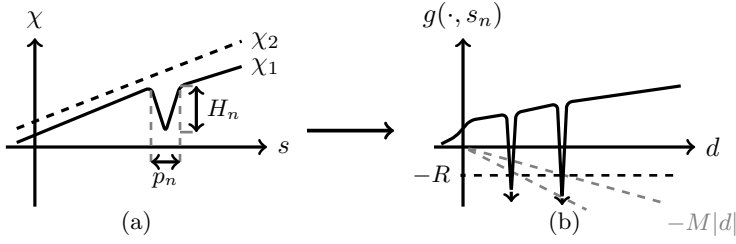

\paragraph{An equivalent working form on the pair}

The regularity package decomposes into two classical geometric conditions on the pair.

\begin{condition}[Bounded closeness]\label{cond:bounded_close}
There exists a constant \(D>0\) such that
\[
\sup_{s\in\R}|\chi_1(s)-\chi_2(s)|\le D.
\]
\end{condition}

\begin{condition}[Large-scale Lipschitzness]\label{cond:large_lip}
There exist constants \(\Lambda>0\) and \(r_0>0\) such that
\[
|\chi_1(x)-\chi_1(y)|\le \Lambda|x-y|,\quad |x-y|\ge r_0.
\]
\end{condition}

\begin{proposition}[Packaging equivalence]\label{prop:packaging}
For a logarithmic comparison pair (finite-valued, \(\chi_1\le\chi_2\)), Condition~\ref{cond:package} is equivalent to the conjunction of Conditions~\ref{cond:bounded_close} and \ref{cond:large_lip}, with explicit constants. If \(|g(d,s)|\le \Lambda|d|\) for all \(|d|\ge r_0\) and all \(s\), then Condition~\ref{cond:bounded_close} holds with \(D=4\Lambda r_0\) and Condition~\ref{cond:large_lip} with constants \((6\Lambda,r_0)\). Conversely, from Condition~\ref{cond:bounded_close} with \(D\) and Condition~\ref{cond:large_lip} with \((\Lambda,r_0)\) one obtains \(|g(d,s)|\le\Lambda|d|+D\) for \(|d|\ge r_0\), and hence Condition~\ref{cond:package} with \(M=\Lambda+2D/r_0\) and any \(R>3\Lambda r_0+D\). Finally, Condition~\ref{cond:package} with constants \((R,M)\) yields Condition~\ref{cond:bounded_close} with \(D=R\) and Condition~\ref{cond:large_lip} with \((M+3R/r_0,r_0)\) for any \(r_0>0\) (the increments satisfy \(|\chi_1(s+u)-\chi_1(s)|=|g(u,s)-g(0,s)|\le M|u|+3R\)).
\end{proposition}

\begin{proof}
We prove the equivalence through the two-sided bound as an intermediate object: the first part relates it to Conditions~\ref{cond:bounded_close} and \ref{cond:large_lip}, and the final paragraph derives it from Condition~\ref{cond:package}.
Assume first the two-sided bound \(|g(d,s)|\le \Lambda|d|\) for \(|d|\ge r_0\). Evaluating at \(d=r_0\) and \(d=2r_0\) with the same \(s\) and subtracting gives \(\chi_1(s+2r_0)-\chi_1(s+r_0)=g(2r_0,s)-g(r_0,s),\) whose absolute value is at most \(3\Lambda r_0\); since every interval \([x,x+r_0]\) is of this form (take \(s=x-r_0\)), the increments of \(\chi_1\) over length \(r_0\) are uniformly bounded by \(3\Lambda r_0\); shorter windows are controlled identically, since \(\chi_1(p)-\chi_1(p-h)=g(r_0+h,p-r_0-h)-g(r_0,p-r_0-h)\) for \(0<h\le r_0\), of absolute value at most \(\Lambda(2r_0+h)\le3\Lambda r_0\). Then \(\chi_2(s)-\chi_1(s)=\bigl[\chi_1(s+r_0)-\chi_1(s)\bigr]-g(r_0,s),\) so \(|\chi_1(s)-\chi_2(s)|\le3\Lambda r_0+\Lambda r_0=4\Lambda r_0\), i.e., Condition~\ref{cond:bounded_close} with \(D=4\Lambda r_0\). For \(|x-y|\ge r_0\), splitting into at most \(2|x-y|/r_0\) segments of length \(r_0\) gives \(|\chi_1(x)-\chi_1(y)|\le6\Lambda|x-y|\), i.e., Condition~\ref{cond:large_lip} with \((6\Lambda,r_0)\). Conversely, if Conditions~\ref{cond:bounded_close} and \ref{cond:large_lip} hold, then for \(|d|\ge r_0\),
\[
|g(d,s)|\le|\chi_1(s+d)-\chi_1(s)|+|\chi_1(s)-\chi_2(s)|
\le\Lambda|d|+D.
\]
For \(|d|<r_0\), transfer through a third point at distance \(r_0\) beyond the far endpoint (\(y=s+r_0+d\) for \(d\ge0\), \(y=s+d-r_0\) for \(d<0\)): writing \(\chi_1(s+d)-\chi_1(s)=[\chi_1(y)-\chi_1(s)]-[\chi_1(y)-\chi_1(s+d)]\), both brackets span distances \(\ge r_0\), so the large-scale Lipschitz bound gives \(|\chi_1(s+d)-\chi_1(s)|\le\Lambda(|d|+r_0)+\Lambda r_0\le3\Lambda r_0\), hence \(|g(d,s)|\le3\Lambda r_0+D\): a dip at a small displacement has uniformly bounded depth. Condition~\ref{cond:package} thus holds with \(M=\Lambda+2D/r_0\) and any \(R>3\Lambda r_0+D\).

It remains to note that the one-sided threshold form \eqref{eq:U1} already yields the two-sided bound with slack. At \(d=0\), the premise \(g(0,s)\le-R\) would force \(g(0,s)\ge0\) by \eqref{eq:U1}, a contradiction; hence \(g(0,s)>-R\), i.e., \(\chi_2(s)-\chi_1(s)\le R\). For arbitrary \((d,s)\), either \(g(d,s)>-R\) or \eqref{eq:U1} applies, so \(g(d,s)\ge-M|d|-R\). This lower bound at \((-d,s+d)\) with \(d>0\) reads \(\chi_1(s)-\chi_2(s+d)\ge-Md-R\), which with \(\chi_2(s+d)\ge\chi_1(s+d)\) controls the rise, \(\chi_1(s+d)-\chi_1(s)\le Md+R\); the lower bound at \((d,s)\) with \(\chi_2(s)\ge\chi_1(s)\) controls the drop, \(\chi_1(s+d)-\chi_1(s)\ge-Md-R\). Combining, \(|g(d,s)|\le|\chi_1(s+d)-\chi_1(s)|+|\chi_1(s)-\chi_2(s)|\le M|d|+2R\). For the increments of \(\chi_1\), note that \(\chi_1(s+u)-\chi_1(s)=g(u,s)-g(0,s)\) with \(|g(u,s)|\le M|u|+2R\) and \(|g(0,s)|\le R\), so \(|\chi_1(s+u)-\chi_1(s)|\le M|u|+3R\); this is Condition~\ref{cond:bounded_close} with \(D=R\) (the \(d=0\) case) and, for \(|u|\ge r_0\), Condition~\ref{cond:large_lip} with \((M+3R/r_0,r_0)\).
\end{proof}

\begin{remark}[Weak same-order as the necessary core]\label{rem:weak_same_order}
In the original coordinates, Condition~\ref{cond:bounded_close} reads \(\alpha_2(r)\le e^{D}\alpha_1(r)\), i.e., a bounded ratio between the two comparison functions---the weak form of the same-order requirement. This weak form is necessary for migration, and it is implied by Hahn's power bounds (with ratio \(\eta_2/\eta_1\)), whereas the power-type requirement itself is not necessary, as shown by He and Wang \cite{he1991}. Thus Hahn's condition splits into a necessary core---a bounded ratio, retained as Condition~\ref{cond:bounded_close}---and an unnecessary power-type strengthening. The remaining two conditions have no counterpart in the classical framework.
\end{remark}

\begin{remark}[Minimality]\label{rem:minimality}
Each part of the condition pair is necessary: the necessity of Conditions~\ref{cond:package} and \ref{cond:joint_lower} is proved directly in Section~\ref{sec:discrete} (Theorem~\ref{thm:strict_beta}), and Proposition~\ref{prop:packaging} then transfers the necessity to Conditions~\ref{cond:bounded_close} and \ref{cond:large_lip} individually.
\end{remark}

We are now ready to state and prove the main result.

\begin{theorem}[Discrete-time strict \(\beta\)-criterion]\label{thm:strict_beta}
Let \((\chi_1,\chi_2)\) be a logarithmic comparison pair. Then the conclusion \eqref{eq:propertyP} of Definition~\ref{def:criterion_logarithm} (with \(t\in\mathbb{N}^*\)) holds if and only if Conditions~\ref{cond:package} and \ref{cond:joint_lower} hold; equivalently, if and only if Conditions~\ref{cond:bounded_close}, \ref{cond:large_lip}, and \ref{cond:joint_lower} hold.
\end{theorem}

\begin{proof}
We first prove sufficiency. Assume that Conditions~\ref{cond:package} (with constants \(R,M\)) and \ref{cond:joint_lower} hold; choose \(d_\beta>0\) such that \(G(d)\ge-\beta\) for all \(d\ge d_\beta\). Note that the package also yields a uniform lower bound on \(G\): for \(d\in[0,d_\beta]\), \(G(d)\ge-Md-R\ge-Md_\beta-R\); together with \(G(d)\ge-\beta\) for \(d\ge d_\beta\), we have \(G_{\min}\ge-H\) with \(H\coloneqq\max\{Md_\beta+R,\,\beta\}\). Set
\[
\eta_\beta\coloneqq\frac{\beta}{M}>0,\qquad
K_{\max}\coloneqq\Bigl\lfloor\frac{H}{\beta}\Bigr\rfloor.
\]
We verify the conclusion in the form \eqref{eq:propertyP_g}. Fix \(s_0\in\R\) and \(k\ge1\), and suppose that \(d\in\R\) satisfies \(g(d,s_0)<-\beta k\).

If \(d\le0\) and \(\beta k\ge R\), then the premise of \eqref{eq:U1} holds, so \(-M|d|\le g(d,s_0)<-\beta k\), i.e., \(M|d|>\beta k\), and hence
\[
d=-|d|<-\frac{\beta k}{M}=-k\eta_\beta\le T_\beta-k\eta_\beta.
\]
If \(\beta k<R\), then \(k<R/\beta\), so \(k\eta_\beta<R/M\), and \(d\le0\le T_\beta-k\eta_\beta\) provided \(T_\beta\ge R/M\).

If instead \(d>0\), then, since \(g(d,s_0)\ge G_{\min}\ge-H\), membership requires \(-\beta k>-H\), i.e., \(k<H/\beta\), and thus \(k\le K_{\max}\) (if \(K_{\max}=0\), the \(d>0\) case is vacuous). If \(d\ge d_\beta\), then \(g(d,s_0)\ge-\beta\ge-\beta k\), contradicting membership; hence \(d<d_\beta\). Therefore \(d<d_\beta\le T_\beta-k\eta_\beta,\) provided \(T_\beta\ge d_\beta+K_{\max}\eta_\beta\).

It remains to unify the constants: taking \(T_\beta\coloneqq\max\left\{\frac{R}{M},\ d_\beta+K_{\max}\frac{\beta}{M}\right\},\) the implication \eqref{eq:propertyP_g} holds for every \(s_0\in\R\) and every \(k\ge1\), and hence \eqref{eq:propertyP} holds. Sufficiency is proved.

For necessity, assume that the conclusion \eqref{eq:propertyP} holds with constants \(T_\beta\ge0\) and \(\eta_\beta>0\). We first prove Condition~\ref{cond:package} by contradiction. Suppose that \eqref{eq:U1} fails: for every \(R>0\) and every \(M'>0\) there exist \(d,s\in\R\) with \(g(d,s)\le-R\) and \(g(d,s)<-M'|d|\). Taking \(R=M'=n\) yields sequences \(\{d_n\}\), \(\{s_n\}\) with \(g(d_n,s_n)\le-n\ (\longrightarrow-\infty),\qquad g(d_n,s_n)<-n|d_n|.\) For all sufficiently large \(n\), set \(k_n\coloneqq\bigl\lceil-g(d_n,s_n)/\beta\bigr\rceil-1\). Since \(\lceil x\rceil-1<x\) for every \(x\), we have \(k_n<-g(d_n,s_n)/\beta\), i.e., \(g(d_n,s_n)<-\beta k_n\). Since \(\lceil x\rceil-1\ge x-2\), we also have \(k_n\ge-g(d_n,s_n)/\beta-2\ge1\) once \(-g(d_n,s_n)\ge3\beta\), i.e., \(-k_n\eta_\beta\le(\eta_\beta/\beta)g(d_n,s_n)+2\eta_\beta\). Then \(d_n\) belongs to the level set \(\{d:g(d,s_n)<-\beta k_n\}\), and the conclusion in the form \eqref{eq:propertyP_g} gives
\[
d_n\le T_\beta-k_n\eta_\beta
\le T_\beta+\frac{\eta_\beta}{\beta}\,g(d_n,s_n)+2\eta_\beta.
\]
If \(d_n\ge0\), this already yields \(-g(d_n,s_n)\le\beta(T_\beta+2\eta_\beta)/\eta_\beta\). If \(d_n<0\), then \(|d_n|<-g(d_n,s_n)/n\), and hence \(\frac{g(d_n,s_n)}{n}<d_n\le T_\beta+\frac{\eta_\beta}{\beta}\,g(d_n,s_n)+2\eta_\beta,\) i.e., \(g(d_n,s_n)\Bigl(\frac1n-\frac{\eta_\beta}{\beta}\Bigr)<T_\beta+2\eta_\beta.\) For \(n>\beta/\eta_\beta\) the bracket is negative, so
\[
g(d_n,s_n)>-\frac{n\beta(T_\beta+2\eta_\beta)}{n\eta_\beta-\beta}\ge-2\beta(T_\beta+2\eta_\beta)/\eta_\beta
\qquad\Bigl(n\ge\frac{2\beta}{\eta_\beta}\Bigr),
\]
contradicting \(g(d_n,s_n)\to-\infty\). Hence \eqref{eq:U1} holds. The constants are tracked directly. Suppose \(g(d,s)\le-R\) with \(R\coloneqq2C\), \(C\coloneqq\beta(T_\beta+2\eta_\beta)/\eta_\beta\), and set \(k\coloneqq\lceil-g(d,s)/\beta\rceil-1\ge1\). Then \(g(d,s)<-\beta k\), so \eqref{eq:propertyP_g} gives \(d\le T_\beta-k\eta_\beta\), while \(-g(d,s)\le\beta(k+1)\); combining, \(-g(d,s)\le(\beta/\eta_\beta)(T_\beta-d)+2\beta\). For \(d\ge0\) this reads \(-g(d,s)\le C\), contradicting \(g(d,s)\le-2C\). For \(d<0\) it reads \(-g(d,s)\le(\beta/\eta_\beta)|d|+C\): if \(|d|\ge C\eta_\beta/\beta\), this is already the desired bound \(-g(d,s)\le(2\beta/\eta_\beta)|d|\); if \(|d|<C\eta_\beta/\beta\), then \(-g(d,s)<2C\), contradicting the premise. Thus \eqref{eq:U1} holds with \(M=2\beta/\eta_\beta\) and \(R=2\beta(T_\beta+2\eta_\beta)/\eta_\beta\).

We now prove Condition~\ref{cond:joint_lower}. Taking \(k=1\) in \eqref{eq:propertyP_g}, whenever \(d>T_\beta-\eta_\beta\), the point \(d\) does not belong to the level set \(\{d:g(d,s_0)<-\beta\}\); otherwise \(d\le T_\beta-\eta_\beta\), a contradiction. Hence \(g(d,s_0)\ge-\beta\) for all \(d>T_\beta-\eta_\beta\). Since \(s_0\) is arbitrary, \(G(d)\ge-\beta\) for all \(d>T_\beta-\eta_\beta\), so Condition~\ref{cond:joint_lower} holds with any \(d_\beta>\max\{T_\beta-\eta_\beta,0\}\).

Necessity is proved, and the working-form statement follows from Proposition~\ref{prop:packaging}.
\end{proof}

\begin{remark}[Explicit rate]\label{rem:rate}
The sufficiency proof is constructive and yields an explicit rate map. If Condition~\ref{cond:package} holds with constants \((R,M)\) and Condition~\ref{cond:joint_lower} holds with threshold \(d_\beta\) (i.e., \(G(d)\ge-\beta\) for \(d\ge d_\beta\)), then the conclusion \eqref{eq:propertyP} holds with \(\eta_\beta=\beta/M\), with the uniform lower bound \(H=\max\{M d_\beta+R,\,\beta\}\) (the same constant as in the proof of Theorem~\ref{thm:strict_beta}), and with
\[
T_\beta=\max\left\{\frac{R}{M},\ d_\beta+\Bigl\lfloor\frac{H}{\beta}\Bigr\rfloor\frac{\beta}{M}\right\}.
\]
In state coordinates this reads \(\norm{x_k}\le C\rho^k\norm{x_0}\) with \(C=e^{T_\beta}\) and \(\rho=e^{-\beta/M}\). Note that the decay rate \(\rho\) depends only on \(\beta\) and the slope constant \(M\) of the package; the threshold \(R\) and the effective depth \(H\) enter only the prefactor \(C\).
\end{remark}

\subsection{The discrete-time general criterion}
\label{subsec:general_criterion}

Theorem~\ref{thm:strict_beta} deals with the discrete-time strict \(\beta\)-criterion for a fixed \(\beta\), with the joint lower bound required at level \(\beta\). For the general criterion, \(\beta\) is no longer fixed; requiring \(G(d)\ge-\beta\) beyond some \(d_\beta\) for every \(\beta>0\) gives, upon intersection over \(\beta\), the new condition \(G_{\lim}\ge0.\) The uniform lower bound \(G_{\min}>-H\) is automatically guaranteed by the regularity package (see Proposition~\ref{prop:packaging} and the proof of Theorem~\ref{thm:strict_beta}) and need not be imposed separately.

\begin{condition}[Joint lower bound (general criterion)]\label{cond:g_general}
With the joint lower envelope \(G\) of \eqref{eq:G}, require
\[
G_{\lim}\coloneqq\liminf_{d\to+\infty}G(d)\ge0.
\]
\end{condition}

\begin{theorem}[Discrete-time general criterion]\label{thm:general}
Let \((\chi_1,\chi_2)\) be a logarithmic comparison pair. Then the discrete-time general criterion holds if and only if Conditions~\ref{cond:package} and \ref{cond:g_general} hold; equivalently, if and only if Conditions~\ref{cond:bounded_close}, \ref{cond:large_lip}, and \ref{cond:g_general} hold.
\end{theorem}

\begin{proof}
For necessity, suppose the general criterion holds. For any \(\beta>0\), the strict \(\beta\)-criterion holds: strictly feasible trajectories are feasible for the general criterion, and at zero initial value the strict feasible set is empty, so the conclusion transfers. By Theorem~\ref{thm:strict_beta}, the pair satisfies Condition~\ref{cond:package} (already at any fixed \(\beta\); the package itself is \(\beta\)-independent), and \(G_{\lim}\ge-\beta\) for every \(\beta>0\). Letting \(\beta\to0^+\), we obtain \(G_{\lim}\ge0\), i.e., Condition~\ref{cond:g_general} holds. The working-form statement follows from Proposition~\ref{prop:packaging}.

For sufficiency, suppose Conditions~\ref{cond:package} and \ref{cond:g_general} hold. For any \(\beta>0\), since \(G_{\lim}\ge0>-\beta\), Condition~\ref{cond:joint_lower} at level \(\beta\) holds. By Theorem~\ref{thm:strict_beta}, the strict \(\beta\)-criterion holds. Since \(\beta\) is arbitrary and the general \(\beta\)-criterion transfers from the strict \(\beta/2\)-criterion (\(\Xi_\beta\subseteq\Xi_{\beta/2}^{\mathrm{s}}\) away from the zero trajectory, whose conclusion is trivial; Section~\ref{subsec:analysis}), the general criterion holds.
\end{proof}

The general criterion differs from the strict \(\beta\)-criterion only in that the level-\(\beta\) threshold is required for every \(\beta>0\), whose intersection is \(G_{\lim}\ge0\); this is necessary, since the plateau mechanism of Figure~\ref{fig:U2_fail} crosses a plateau at a fixed gap \(h\), and validity for arbitrarily small \(\beta\) forces \(h\to0\).

\begin{figure}[tb]
\centering
\begin{tikzpicture}[scale=0.48,yscale=0.82]
\begin{scope}
    \draw[->,very thick] (-0.8,0) -- (6.4,0) node[right] {\(s\)};
    \draw[->,very thick] (0,-2.0) -- (0,3.65) node[above] {\(\chi\)};
    \draw[very thick,cblue,rounded corners=2pt] (-0.5,-0.05) -- (0.3,0.5) -- (1.1,1.0) -- (2.2,1.0) -- (2.9,1.7) -- (4.4,1.7) -- (5.0,2.1) -- (5.8,2.1);
    \node[below right] at (5.8,2.1) {\(\chi_1\)};
    \draw[very thick,black,dashed,rounded corners=2pt] (-0.5,0.25) -- (0.3,0.7) -- (1.1,2.0) -- (2.2,2.0) -- (2.9,2.4) -- (4.4,2.4) -- (5.0,2.6) -- (5.8,2.6);
    \node[above right] at (5.8,2.6) {\(\chi_2\)};
    \draw[<->,very thick] (1.65,1.0) -- (1.65,2.0) node[above]{\small \(h_1\)};
    \draw[<->,very thick] (3.65,1.7) -- (3.65,2.4) node[above]{\small \(h_2\)};
    \node at (2.8,-2.55) {\small (a)};
\end{scope}
\draw[->,line width=1.4pt] (7.5,0.55) -- (9.9,0.55);
\begin{scope}[shift={(11.8,0)}]
    \draw[->,very thick] (-0.8,0) -- (6.4,0) node[right] {\(d\)};
    \draw[->,very thick] (0,-2.0) -- (0,3.65);
    \node[above] at (0,3.65) {\(g(\cdot,a_j)\)};
    \draw[very thick,cblue,rounded corners=4pt] (-0.6,0.6) -- (-0.2,0.78) -- (0.2,0.85) -- (0.55,0.7) -- (0.8,-0.9) -- (5.15,-0.9) -- (5.5,0.7) -- (6.05,0.95);
    \draw[<->,very thick] (1.25,-0.9) -- (1.25,0) node[above]{\small \(h_j\)};
    \draw[<->,very thick] (0.8,-1.2) -- (5.15,-1.2) node[right]{\small \(b_j-a_j\to\infty\)};
    \node at (2.8,-2.55) {\small (b)};
\end{scope}
\end{tikzpicture}
\caption{The discrete-time general criterion (Condition~\ref{cond:g_general}). (a) Plateaus of unbounded length are admissible provided their depths \(h_j\) tend to zero. (b) On the displacement function, each plateau produces a trench of depth \(h_j\) and unbounded width, so \(G(d)\) may take negative values as long as \(G(d)\to0^{-}\).}
\label{fig:general_discrete}
\end{figure}
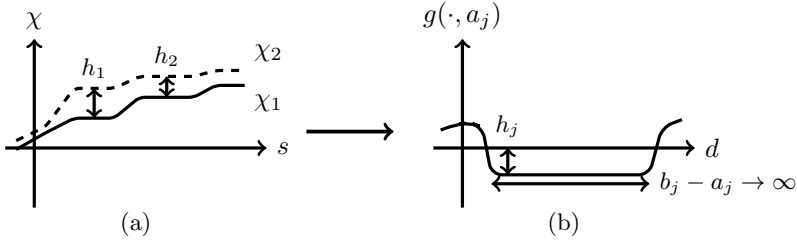

Plateau lengths that grow without bound are admissible in discrete time provided that their depths tend to zero; see Figure~\ref{fig:general_discrete}.

The discrete-time general criterion transfers to the weakened continuous-time one without further work:

\begin{corollary}[Weakened continuous-time criterion]\label{cor:general}
Let \((\chi_1,\chi_2)\) be a logarithmic comparison pair. Then the discrete-time general criterion holds if and only if Conditions~\ref{cond:package} and \ref{cond:g_general} hold. Furthermore, the weakened continuous-time general criterion (Definition~\ref{def:criterion_weak}) holds if and only if the same conditions hold.
\end{corollary}

\begin{proof}
By Theorem~\ref{thm:cont_disc}, the weakened continuous-time general criterion is equivalent to the discrete-time general criterion, so their necessary and sufficient conditions are the same.
\end{proof}

\begin{remark}[A variable-power pair]\label{rem:varpower}
The conditions of Theorem~\ref{thm:general} require neither fixed powers nor fixed slopes. A piecewise power function whose powers cycle through \(1,2,4,\tfrac12\) over intervals of exponentially growing lengths yields \(\chi_2=\chi_1\) with slopes in \([\tfrac12,4]\), so \(g(d,s)\ge0\) identically for \(d\ge0\) and the conditions hold trivially, while Hahn's same-power bounds fail. \ifextended The complete construction and verification are given in Appendix~\ref{app:varpower}.\else The complete construction and verification are given in Appendix~A of the extended version~\cite{guo2026extended}.\fi
\end{remark}

\paragraph{A fast-decay pair}\label{subsec:ex_fastdecay}

We conclude this section with a pair that realizes the separation between the discrete-time and continuous-time criteria: the discrete-time conditions hold while the continuous-time ones fail, and the pair lies beyond every fixed-order criterion.

\[
\alpha_1(r)=\ln(\ln r),\qquad\alpha_2(r)=\ln(2\ln r),\quad r>e,
\]
extended by tangent lines from the origin: the tangency point \(r_\star>e\) for \(\alpha_1\) solves \(\alpha_1(r_\star)=r_\star\alpha_1'(r_\star)\), i.e., \(u\ln u=1\) with \(u=\ln r_\star\), which has a unique solution \(u_1\in(1,e)\); one keeps \(\ln(\ln r)\) beyond \(r_\star\) and sets \(\alpha_1(r)=\alpha_1(r_\star)r/r_\star\) for \(r\le r_\star\), and the same procedure applies to \(\alpha_2\). The tangent condition matches the slopes at the junction, so the extended \(\alpha_i\) are \(C^1\) and strictly increasing on \(\Rnn\) with \(\alpha_i(0)=0\), and \(\chi_1'\le1\) throughout. The ordering \(\alpha_1\le\alpha_2\) is inherited: the tangent slopes \(e^{-u_i}/u_i\) are ordered by \(u_2<u_1\) (the tangency equations are \(u\ln u=1\) and \(u\ln(2u)=1\)), and on the mixed segment \(\alpha_1(r)\le\ln u_1<\ln(2u_2)\le\alpha_2(r)\) (indeed \(u_1<2u_2\): \(u_1\ge2u_2\) would give \(1=u_1\ln u_1\ge2u_2\ln(2u_2)=2\)). The effective power \(\chi_1(s)/s\to0\) as \(s\to\infty\) (recall \(s=\ln r\)), so no fixed-order bounds apply. Bounded closeness holds since
\[
\chi_2(s)-\chi_1(s)=\ln\frac{\ln(2s)}{\ln s}=\ln\!\left(1+\frac{\ln 2}{\ln s}\right)
\]
is bounded, and \(\chi_1\) is globally Lipschitz, so Condition~\ref{cond:large_lip} holds. For the joint lower bound,
\[
g(d,s)=\chi_1(s+d)-\chi_2(s)=\ln\frac{\ln(s+d)}{\ln(2s)},
\]
so \(g(d,s)\ge0\) if and only if \(d\ge s\). Hence, for every fixed \(d>0\), choosing \(s>d\) gives \(g(d,s)<0\): \(G(d)<0\) for every \(d>0\), the set \(\{d\ge0:G(d)<0\}\) is unbounded, Condition~\ref{cond:g_continuous} fails, and by Theorem~\ref{thm:continuous} (Section~\ref{sec:continuous}) the continuous-time general criterion does not hold for this pair. On the other hand, since \(\chi_1\) is nondecreasing, \(g(d,s)\ge\chi_1(s)-\chi_2(s)\ge-D\), and, for \(s>d\),
\[
g(d,s)\ge\ln\frac{\ln s}{\ln(2s)}\ge\ln\frac{\ln d}{\ln d+\ln2}\longrightarrow0^{-}\quad(d\to+\infty),
\]
so \(G_{\min}\ge-D>-\infty\) and \(G_{\lim}=0\): the three conditions of Theorem~\ref{thm:general} hold, and the discrete-time general criterion covers this pair. This example therefore realizes the separation of Theorem~\ref{thm:continuous} (Section~\ref{sec:continuous}): the discrete-time criterion accepts what the continuous-time criterion rejects.

\begin{remark}[A closed-form skeleton pair]\label{rem:trench_rate}
One can go further with a closed-form pair quantifying the approach to the continuous-time boundary. With \(a_j=2e^j\), \(b_j=a_{j+1}-1+\varepsilon\sin j\) (\(\varepsilon=0.3\)), \(\eta_0=a_0\), and \(\eta_{j+1}=\eta_j-e^{-j^2}(b_j-a_j)+2(a_{j+1}-b_j)\), define the skeleton
\[
\chi_0(s)=
\begin{cases}
s, & s\le a_0,\\[3pt]
\eta_j-e^{-j^2}(s-a_j), & s\in[a_j,b_j],\\[3pt]
\eta_j-e^{-j^2}(b_j-a_j)+2(s-b_j), & s\in[b_j,a_{j+1}],
\end{cases}
\]
and set \(\chi_1=\chi_0-\bar u\), \(\chi_2=\chi_0+\bar u\), where \(\bar u(s)=A\) for \(s\le a_0\) and \(\bar u\) interpolates \(A/(j+1)^2\) and \(A/(j+2)^2\) linearly on \((a_j,a_{j+1}]\) (\(A=0.15\)). The resulting pair satisfies \(G(d)<0\) at every \(d>0\), \(G_{\lim}=0\), and \(-G(d)\sim2A(\ln d)^{-2}\) as \(d\to\infty\). \ifextended The analysis is given in Appendix~\ref{app:skeleton}.\else The analysis is given in Appendix~B of the extended version~\cite{guo2026extended}.\fi
\end{remark}

\section{The continuous-time criterion}
\label{sec:continuous}

We now turn to continuous time. By Corollary~\ref{cor:general}, the weakened continuous-time general criterion is already settled by the discrete-time one: for \(t\ge\varepsilon\) the two settings coincide, and the joint lower bound is \(G_{\lim}\ge0\). What remains is the initial segment \(t\in(0,\varepsilon)\). There the elapsed time can be arbitrarily small, so no minimum decay per step is available: if \(G_{\lim}=0\) while \(G(d)<0\) at arbitrarily large \(d\), then for arbitrarily small \(t\) the level-set condition \(g(d,s_0)<-\beta t\) can still be met with \(d\) unbounded, escaping the control of \(T_\beta-t\eta_\beta\).

\begin{condition}[Strengthened joint lower bound]\label{cond:g_continuous}
With the joint lower envelope \(G\) of \eqref{eq:G}, require that there exists \(d_{\max}\ge0\) such that for every \(d>d_{\max}\),
\[
G(d)\ge0.
\]
\end{condition}

Condition~\ref{cond:g_continuous} is equivalent to requiring the set \(\{d\ge0:G(d)<0\}\) to be bounded; it implies \(G_{\lim}\ge0\) but not conversely, since \(G_{\lim}\ge0\) tolerates small negative values of \(G\) at arbitrarily large \(d\).

\begin{figure}[tb]
\centering
\begin{tikzpicture}[scale=0.48,yscale=0.82]
\begin{scope}
    \draw[->,very thick] (-0.8,0) -- (6.4,0) node[right] {\(s\)};
    \draw[->,very thick] (0,-2.0) -- (0,3.65) node[above] {\(\chi\)};
    \draw[very thick,cblue,rounded corners=3pt] (-0.5,0.24) -- (0.4,0.75) -- (1.9,1.6) -- (3.0,1.6) -- (5.8,2.9);
    \node[right] at (5.8,2.9) {\(\chi_1\)};
    \draw[very thick,black,dashed,rounded corners=3pt] (-0.5,0.38) -- (0.4,1.1) -- (1.9,2.45) -- (3.0,2.45) -- (5.8,3.55);
    \node[right] at (5.8,3.55) {\(\chi_2\)};
    \draw[dashed,gray,line width=1pt] (1.9,-0.7) -- (1.9,1.6);
    \draw[dashed,gray,line width=1pt] (3.0,-0.7) -- (3.0,1.6);
    \node[anchor=east] at (1.75,-0.4) {\(a\)};
    \node[anchor=west] at (3.15,-0.4) {\(b\)};
    \draw[<->,very thick] (2.45,1.6) -- (2.45,2.45) node[above]{\small \(h\)};
    \draw[<->,very thick] (1.9,-0.7) -- (3.0,-0.7) node[midway,below]{\small \(b-a\)};
    \node at (2.8,-2.55) {\small (a)};
\end{scope}
\draw[->,line width=1.4pt] (7.5,0.55) -- (9.9,0.55);
\begin{scope}[shift={(11.8,0)}]
    \draw[->,very thick] (-0.8,0) -- (6.4,0) node[right] {\(d\)};
    \draw[->,very thick] (0,-2.0) -- (0,3.65);
    \node[above] at (0,3.65) {\(g(\cdot,a)\)};
    \draw[very thick,cblue,rounded corners=4pt] (-0.6,0.5) -- (-0.2,0.65) -- (0.2,0.7) -- (0.55,0.55) -- (0.8,-0.95) -- (2.7,-0.95) -- (3.1,0.6) -- (6.0,1.3);
    \draw[<->,very thick] (1.3,-0.95) -- (1.3,0) node[above]{\small \(h\)};
    \draw[<->,very thick] (0.8,-1.45) -- (2.7,-1.45) node[right]{\small \(\le d_{\max}\)};
    \draw[dashed,gray,line width=1pt] (3.65,-0.95) -- (3.65,0.5);
    \node[below right] at (3.65,0.02) {\small \(d_{\max}\)};
    \node at (2.8,-2.55) {\small (b)};
\end{scope}
\end{tikzpicture}
\caption{The strengthened joint lower bound in continuous time (Condition~\ref{cond:g_continuous}). (a) A plateau of fixed depth \(h\) is admissible only if its length is bounded. (b) Equivalently, the trench of \(g(\cdot,a)\) must terminate: negative values of \(G\) are confined to a bounded interval \([0,d_{\max}]\).}
\label{fig:continuous_strengthened}
\end{figure}
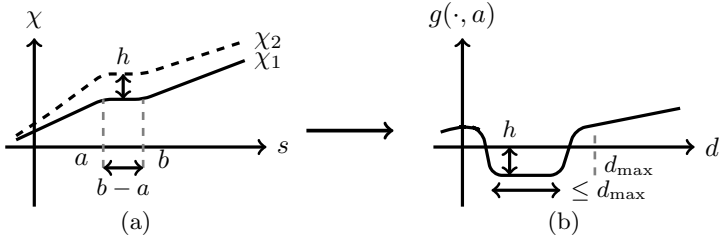

Figure~\ref{fig:continuous_strengthened} contrasts with Figure~\ref{fig:general_discrete}: in continuous time the trench must terminate.

\begin{theorem}[Continuous-time general criterion]\label{thm:continuous}
Let \((\chi_1,\chi_2)\) be a logarithmic comparison pair. Then the continuous-time general criterion holds if and only if Conditions~\ref{cond:package} and \ref{cond:g_continuous} hold; equivalently, if and only if Conditions~\ref{cond:bounded_close}, \ref{cond:large_lip}, and \ref{cond:g_continuous} hold.
\end{theorem}

\begin{proof}
For sufficiency, assume that Conditions~\ref{cond:package} and \ref{cond:g_continuous} hold with threshold \(d_{\max}\ge0\), i.e., \(G(d)\ge0\) for \(d>d_{\max}\). Fix \(\varepsilon>0\). By Corollary~\ref{cor:general}, the weakened continuous-time general criterion holds, i.e., there exist \(T_{\beta,\varepsilon}\ge0\) and \(\eta_{\beta,\varepsilon}>0\) such that for every \(s_0\in\R\) and every \(t\ge\varepsilon\), \(\Gamma_\beta(s_0,t)\le s_0+T_{\beta,\varepsilon}-t\eta_{\beta,\varepsilon}.\) We now show that the conclusion also holds for \(t\in(0,\varepsilon)\). Fix \(\beta>0\). If \(\Gamma_\beta(s_0,t)>s_0+d_{\max}\), then there exists \(s>s_0+d_{\max}\) such that \(\chi_1(s)<\chi_2(s_0)-\beta t\), i.e., \(g(s-s_0,s_0)<0\). This contradicts Condition~\ref{cond:g_continuous}, since \(s-s_0>d_{\max}\). Hence \(\Gamma_\beta(s_0,t)\le s_0+d_{\max},\quad t\in(0,\varepsilon).\) Take \(T_\beta\coloneqq\max\{T_{\beta,\varepsilon},\ d_{\max}+\varepsilon\eta_{\beta,\varepsilon}\}\). For \(t\in(0,\varepsilon)\),
\[
\Gamma_\beta(s_0,t)\le s_0+d_{\max}\le s_0+T_\beta-\varepsilon\eta_{\beta,\varepsilon}\le s_0+T_\beta-t\eta_{\beta,\varepsilon},
\]
where the middle step uses \(T_\beta\ge d_{\max}+\varepsilon\eta_{\beta,\varepsilon}\) and the last step uses \(t<\varepsilon\). For \(t\ge\varepsilon\), since \(T_\beta\ge T_{\beta,\varepsilon}\),
\[
\Gamma_\beta(s_0,t)\le s_0+T_{\beta,\varepsilon}-t\eta_{\beta,\varepsilon}\le s_0+T_\beta-t\eta_{\beta,\varepsilon}.
\]
Therefore, for every \(t>0\), \(\Gamma_\beta(s_0,t)\le s_0+T_\beta-t\eta_\beta\) holds with \(\eta_\beta\coloneqq\eta_{\beta,\varepsilon}\), and sufficiency for the continuous-time general criterion follows.

For necessity, assume that the continuous-time general criterion holds, i.e., for every \(\beta>0\), there exist \(T_\beta\ge0\) and \(\eta_\beta>0\) such that for every \(s_0\in\R\) and every \(t>0\), \(\Gamma_\beta(s_0,t)\le s_0+T_\beta-t\eta_\beta.\) The necessity of Condition~\ref{cond:package} follows directly from Corollary~\ref{cor:general} and Theorem~\ref{thm:strict_beta}. We now prove Condition~\ref{cond:g_continuous}.

Suppose there exists a sequence \(d_n\to+\infty\) such that \(G(d_n)<0\). Take \(\beta>0\) and \(t_n>0\) such that \(\beta t_n<|G(d_n)|\) (note \(|G(d_n)|>0\), since \(G(d_n)<0\)). By the definition of \(G(d_n)=\inf_s g(d_n,s)\), there exists \(s_n\in\R\) such that \(g(d_n,s_n)<-\beta t_n\), i.e., \(\chi_1(s_n+d_n)<\chi_2(s_n)-\beta t_n\). Hence \(s_n+d_n\) belongs to the level set, and thus \(\Gamma_\beta(s_n,t_n)\ge s_n+d_n.\) By \eqref{eq:propertyP}, \(\Gamma_\beta(s_n,t_n)\le s_n+T_\beta-t_n\eta_\beta\), so \(d_n\le T_\beta-t_n\eta_\beta<T_\beta\). Since \(d_n\to+\infty\), this contradicts the finiteness of \(T_\beta\). Therefore, there exists \(d_{\max}\ge0\) such that \(G(d)\ge0\) for \(d>d_{\max}\). Condition~\ref{cond:g_continuous} holds.

\end{proof}

The continuous-time criterion thus differs from the discrete-time one only in the strengthening of Condition~\ref{cond:g_general} to \ref{cond:g_continuous}.

\subsection{Relation to the Panteley--Lor\'ia criterion}
\label{subsec:panteley}

Let \(\alpha_1,\alpha_2\in\mathcal{K}_\infty\). The Panteley--Lor\'ia criterion \cite{panteley2010} requires that there exist constants \(c_1>0\), \(c_2>0\), and \(c_3\in(0,1)\) such that
\begin{equation}\label{eq:panteley_conditions}
\alpha_1^{-1}(\alpha_2(r))\le c_1 r,\qquad
\alpha_1^{-1}(c_2\alpha_2(r))\le c_3 r,\quad\forall r\ge0.
\end{equation}
(The inner scaling constant is required only to be positive in \cite{panteley2010}; under the strict monotonicity of \(\alpha_1\), the second inequality already forces \(c_2<1\): in logarithmic coordinates it reads \(\chi_2(s)+\ln c_2\le\chi_1(s+\ln c_3)\), and since \(\ln c_3<0\) and \(\chi_1\) is strictly increasing, \(\chi_1(s+\ln c_3)<\chi_1(s)\le\chi_2(s)\), which would contradict \(c_2\ge1\).)

The inverse function in \eqref{eq:panteley_conditions} is only notation: for \(\alpha_1\) nondecreasing, \(\alpha_1^{-1}(u)\le t\) if and only if \(u\le\alpha_1(t)\). Eliminating the inverse yields the conditions in substantive form.

\begin{condition}[Direct Panteley--Lor\'ia conditions]\label{cond:pl_direct}
There exist constants \(c_1>0\), \(c_2>0\), and \(c_3\in(0,1)\) such that, for all \(r\ge0\),
\begin{equation}\label{eq:pl_direct}
\alpha_2(r)\le\alpha_1(c_1 r),\qquad
c_2\alpha_2(r)\le\alpha_1(c_3 r).
\end{equation}
\end{condition}

Unlike \eqref{eq:panteley_conditions}, the direct form \eqref{eq:pl_direct} is meaningful for arbitrary pairs. In logarithmic coordinates the two inequalities read
\begin{equation}\label{eq:pl_log}
\chi_2(s)\le\chi_1(s+\ln c_1),\qquad
\chi_2(s)-\gamma\le\chi_1(s-c),\qquad \gamma\coloneqq-\ln c_2,\ c\coloneqq-\ln c_3>0.
\end{equation}

\begin{proposition}[The direct Panteley--Lor\'ia conditions are ours on monotone pairs]\label{prop:pl_equiv}
For a logarithmic comparison pair with \(\chi_1\) nondecreasing, Condition~\ref{cond:pl_direct} holds if and only if Conditions~\ref{cond:package} and \ref{cond:g_continuous} hold together, with explicit constants: given \eqref{eq:pl_direct}, Condition~\ref{cond:g_continuous} holds with \(d_{\max}=\max\{\ln c_1,0\}\), and the pair satisfies Conditions~\ref{cond:bounded_close} and \ref{cond:large_lip} with \(D=\gamma\lceil\max\{\ln c_1,0\}/c\rceil\) (or any positive value should this quantity vanish) and \((\Lambda,r_0)=(2\gamma/c,c)\), hence Condition~\ref{cond:package} by Proposition~\ref{prop:packaging}; conversely, \eqref{eq:pl_direct} holds with any \(c_1>e^{d_{\max}}\), \(c_3=e^{-r_0}\), and \(c_2=e^{-(D+\Lambda r_0)}\).
\end{proposition}

\begin{proof}
In the form \eqref{eq:pl_log}, the first inequality reads \(\chi_2(s)\le\chi_1(s+\ln c_1)\) for all \(s\); since \(\chi_1\) is nondecreasing, \(\chi_1(s+d)\ge\chi_2(s)\) for every \(d\ge\ln c_1\)---in particular for every \(d>0\) when \(\ln c_1<0\)---which is Condition~\ref{cond:g_continuous} with \(d_{\max}=\max\{\ln c_1,0\}\). Conversely, \(G(d)\ge0\) for \(d>d_{\max}\) is the first direct inequality with any \(c_1>e^{d_{\max}}\). For the rest, recall from Proposition~\ref{prop:packaging} that Condition~\ref{cond:package} is Conditions~\ref{cond:bounded_close} and \ref{cond:large_lip} together. The second direct inequality reads \(\chi_1(t-c)\ge\chi_2(t)-\gamma\ge\chi_1(t)-\gamma\), i.e., the rise of \(\chi_1\) over every window of length \(c=-\ln c_3\) is at most \(\gamma\); subdividing a window of length \(L\ge c\) into \(\lceil L/c\rceil\le2L/c\) such windows gives Condition~\ref{cond:large_lip} with \((2\gamma/c,c)\). Chaining with the first inequality gives \(\chi_2(s)\le\chi_1(s+\max\{\ln c_1,0\})\le\chi_1(s)+\gamma\lceil\max\{\ln c_1,0\}/c\rceil\), Condition~\ref{cond:bounded_close} with \(D=\gamma\lceil\max\{\ln c_1,0\}/c\rceil\). The edge cases are covered as follows: if \(\ln c_1<0\), monotonicity alone gives \(\chi_2\le\chi_1\); if \(0<\max\{\ln c_1,0\}<c\), then \(\lceil\max\{\ln c_1,0\}/c\rceil=1\) and the rise over the shorter window is still at most \(\gamma\); and since Condition~\ref{cond:bounded_close} requires \(D>0\), should the formula return \(D=0\), any positive number serves. Conversely, large-scale Lipschitzness and bounded closeness give \(\chi_1(s-c)\ge\chi_2(s)-(D+\Lambda c)\), the second direct inequality with \(\gamma=D+\Lambda c\) and \(c=r_0\).
\end{proof}

\begin{remark}[Monotonicity cannot be dropped]\label{rem:pl_sharp}
The monotonicity hypothesis of Proposition~\ref{prop:pl_equiv} is essential, even among continuous pairs with both tails proper. Take \(\chi_1=\chi_2=\ln s+A\sin(2\pi s/a)\) for \(s\ge1\), extended to \(s<1\) by a slope-1 ray, so that \(\chi_1(\pm\infty)=\pm\infty\), taking \(0<A\le\tfrac12\ln(1+a)\). The first direct inequality holds with \(c_1=e^a\): for \(s\ge1\), \(\chi_1(s+a)-\chi_1(s)=\ln(1+a/s)\ge0\) (the sine cancels); on the ray \(s<1-a\) the increment is \(a>0\); and on the mixed segment \(s\in[1-a,1)\) it equals \(\ln(s+a)-(s-1)+A[\sin(2\pi s/a)-\sin(2\pi/a)]\ge\ln(1+a)-2A\ge0\); the second holds with \(\gamma=c+2A\), since the window variation of \(\chi_1\) is at most \(c+2A\). Yet the dip-aligned choice \(d_n=k_na+a/2\), \(s_n=na+a/4\) with \(k_n=\lfloor\sqrt n\rfloor\) gives \(d_n\to\infty\) and \(d_n/s_n\to0\), and \(g(d_n,s_n)=\ln(1+d_n/s_n)-2A\to-2A\), so \(G(d_n)<-A\) at arbitrarily large \(d_n\): by Theorems~\ref{thm:general} and \ref{thm:continuous}, neither criterion holds for this pair.
\end{remark}

\begin{remark}[The Panteley--Lor\'ia criterion is complete on the classical class once the inverse is removed]\label{rem:pl_complete}
The inverse-function formulation is what imports the \(\mathcal{K}_\infty\) regularity---continuity, strict monotonicity, radial unboundedness---into the statement of \cite{panteley2010}: without the inverse, the criterion cannot be written, since \(\alpha_1^{-1}\) exists as a function only in that class. Proposition~\ref{prop:pl_equiv} locates the content of that regularity precisely: written directly, the Panteley--Lor\'ia criterion coincides with our necessary and sufficient continuous-time conditions on the nondecreasing pairs---a strict widening of the classical class (and monotonicity cannot be relaxed further; Remark~\ref{rem:pl_sharp}). Moreover, every admissible pair reaches the nondecreasing class by monotonizing its lower bound (Remark~\ref{rem:pl_reachable}). In this sense the criterion of \cite{panteley2010} was already complete on the classical regularity class; this coincidence does not seem to have been observed before. First, the ``necessity'' in \cite{panteley2010} is read existentially---via a converse theorem producing \emph{some} quadratic-bounded Lyapunov function obeying the growth conditions---and is therefore invisible at the level of the prescribed pair; criteria over prescribed pairs were not yet objects of study. Second, the gap between the classical class and the nondecreasing class is genuine: requiring the inverse, the criterion of \cite{panteley2010} reaches neither the sawtooth pair of Section~\ref{subsec:ex_sawtooth}, whose \(\alpha_1\) is not monotone, nor the pair with plateaus of unbounded length \ifextended of Appendix~\ref{app:strict}\else constructed in \cite[Appendix~C]{guo2026extended}\fi, which admits no strictly increasing refinement at all---while the direct form \eqref{eq:pl_direct} remains meaningful and decisive on both. In discrete time, finally, our condition \(G_{\lim}\ge0\) is strictly weaker than \eqref{eq:pl_direct} already on the nondecreasing class: the fast-decay pair of Section~\ref{subsec:ex_fastdecay}, whose \(\chi_1\) is increasing, satisfies our discrete-time conditions while violating the first inequality of \eqref{eq:pl_direct}.
\end{remark}

\section{Envelopes and verification of the conditions}
\label{sec:envelope}

The conditions of Section~\ref{sec:conditions} allow the comparison pair to be discontinuous, non-monotone, or non-Lipschitz. This section shows that nothing is lost by regularization: every admissible pair admits smooth, nondecreasing, Lipschitz envelopes with the same discriminative power---a canonical procedure for polishing any admissible pair into regular form.

\subsection{Envelope construction}
\label{subsec:envelope_construction}

Let \((\chi_1,\chi_2)\) satisfy Conditions~\ref{cond:bounded_close}, \ref{cond:large_lip}, and \ref{cond:g_general} (discrete-time) or \ref{cond:g_continuous} (continuous-time). We construct smooth, nondecreasing, Lipschitz envelopes \(L,U\) in four steps such that \(L\le\chi_1\le\chi_2\le U,\) and property \eqref{eq:propertyP} is preserved. Throughout this section the pair is understood to arise from a Lyapunov function, so the left tail \(\chi_2(-\infty)=-\infty\) is in force (Section~\ref{subsec:tails}).

For any logarithmic comparison pair satisfying property \eqref{eq:propertyP}, Theorem~\ref{thm:strict_beta} implies that it satisfies the three conditions, possibly with different constants (which may be read off at any fixed \(\beta\), e.g., \(\beta=1\)); we denote the resulting Lipschitz constant and large-scale threshold of the semi-continuous stage by \(\ell_{\mathrm{sc}}\) and \(r_{\mathrm{sc}}\).

We first monotonize the pair: define
\[
L_{\mathrm{mon}}(s)\coloneqq\inf_{u\ge s}\chi_1(u),\qquad
U_{\mathrm{mon}}(s)\coloneqq\sup_{u\le s}\chi_2(u).
\]
The uniform lower bound holds with an explicit constant: \(G_{\lim}\ge0\) bounds \(G\) below at all large \(d\), and the regularity package gives \(G(d)\ge-Md-R\) on the remaining bounded interval, so \(G_{\min}\ge-H\) for some \(H>0\). Then \(L_{\mathrm{mon}}(s)=\inf_{u\ge s}\chi_1(u)\ge\chi_2(s)-H\), so \(L_{\mathrm{mon}}\) is finite, nondecreasing, and \(L_{\mathrm{mon}}\le\chi_1\); and for \(u\le s\) the same bound gives \(\chi_2(u)\le\chi_1(s)+H\), so \(U_{\mathrm{mon}}(s)=\sup_{u\le s}\chi_2(u)\) is finite, nondecreasing, and satisfies \(U_{\mathrm{mon}}\ge\chi_2\). (The standing left tail is thus not what makes the envelopes finite; it is assumed for the correspondence with a continuous Lyapunov function.)

Since \(L_{\mathrm{mon}}(s)<c\) holds if and only if \(\chi_1(u)<c\) for some \(u\ge s\), taking suprema yields
\begin{equation}\label{eq:mono_invariance}
\sup\{s:L_{\mathrm{mon}}(s)<c\}=\sup\{s:\chi_1(s)<c\}.
\end{equation}

Fix \(s_0\in\R\) and an integer \(t\ge1\) (only integer times are needed, since the envelope is used in the discrete-time framework). By the definition of \(\Gamma^{L_{\mathrm{mon}},U_{\mathrm{mon}}}\) and \eqref{eq:mono_invariance} with \(c=U_{\mathrm{mon}}(s_0)-\beta t\), \(\Gamma^{L_{\mathrm{mon}},U_{\mathrm{mon}}}(s_0,t)=\sup\{s:\chi_1(s)<U_{\mathrm{mon}}(s_0)-\beta t\}.\) Since \(U_{\mathrm{mon}}(s_0)=\sup_{u\le s_0}\chi_2(u)\), the strict inequality \(\chi_1(s)<U_{\mathrm{mon}}(s_0)-\beta t\) holds if and only if \(\chi_1(s)<\chi_2(u)-\beta t\) for some \(u\le s_0\). Hence
\[\Gamma^{L_{\mathrm{mon}},U_{\mathrm{mon}}}(s_0,t)=\sup_{u\le s_0}\Gamma^\chi(u,t)
\le\sup_{u\le s_0}\bigl(u+T_\beta-t\eta_\beta\bigr)\le s_0+T_\beta-t\eta_\beta,\]
where the first inequality applies property \eqref{eq:propertyP} at each starting point \(u\).

Next, we make the pair semi-continuous: \(L_{\mathrm{sc}}(s)\coloneqq L_{\mathrm{mon}}(s^-)\le L_{\mathrm{mon}}(s)\) and \(U_{\mathrm{sc}}(s)\coloneqq U_{\mathrm{mon}}(s^+)\ge U_{\mathrm{mon}}(s)\) (the one-sided limits exist by monotonicity). If \(s>\sup\{u:L_{\mathrm{mon}}(u)<c\}\), then \(L_{\mathrm{mon}}\ge c\) on the interval in between, so \(L_{\mathrm{sc}}(s)=\sup_{u<s}L_{\mathrm{mon}}(u)\ge c\); together with \(L_{\mathrm{sc}}\le L_{\mathrm{mon}}\), the two level sets have the same supremum:
\begin{equation}\label{eq:cont_invariance}
\sup\{s:L_{\mathrm{sc}}(s)<c\}=\sup\{s:L_{\mathrm{mon}}(s)<c\}.
\end{equation}

Since \(U_{\mathrm{sc}}(s_0)=U_{\mathrm{mon}}(s_0^+)\le U_{\mathrm{mon}}(s_0+\delta)\) for every \(\delta>0\), the level-set invariance \eqref{eq:cont_invariance} with \(c=U_{\mathrm{sc}}(s_0)-\beta t\) and the monotonicity of the level-set supremum in the threshold give
\[
\begin{aligned}
\Gamma^{L_{\mathrm{sc}},U_{\mathrm{sc}}}(s_0,t)&=\sup\{s:L_{\mathrm{sc}}(s)<c\}=\sup\{s:L_{\mathrm{mon}}(s)<c\}\\
&\le\sup\{s:L_{\mathrm{mon}}(s)<U_{\mathrm{mon}}(s_0+\delta)-\beta t\}\\
&=\Gamma^{L_{\mathrm{mon}},U_{\mathrm{mon}}}(s_0+\delta,t)\le s_0+\delta+T_\beta-t\eta_\beta.
\end{aligned}
\]
Letting \(\delta\to0^+\) yields \(\Gamma^{L_{\mathrm{sc}},U_{\mathrm{sc}}}(s_0,t)\le s_0+T_\beta-t\eta_\beta\).

The pair \((L_{\mathrm{sc}},U_{\mathrm{sc}})\) is again a logarithmic comparison pair, and the left tail is inherited from \(\chi_2\). Since it satisfies property \eqref{eq:propertyP}, Theorem~\ref{thm:strict_beta} implies that it satisfies the three conditions on the pair, with Lipschitz constant \(\ell_{\mathrm{sc}}\), large-scale threshold \(r_{\mathrm{sc}}\), and closeness constant \(D\), i.e., \(U_{\mathrm{sc}}-L_{\mathrm{sc}}\le D\). The large-scale Lipschitzness of \(U_{\mathrm{sc}}\) then follows from that of \(L_{\mathrm{sc}}\): by the triangle inequality,
\[\begin{aligned}|U_{\mathrm{sc}}(x)-U_{\mathrm{sc}}(y)|&\le|U_{\mathrm{sc}}(x)-L_{\mathrm{sc}}(x)|+|L_{\mathrm{sc}}(x)-L_{\mathrm{sc}}(y)|+|L_{\mathrm{sc}}(y)-U_{\mathrm{sc}}(y)|\\
&\le2D+\ell_{\mathrm{sc}}|x-y|,\end{aligned}\]
so for \(|x-y|\ge r_{\mathrm{sc}}\), the increment of \(U_{\mathrm{sc}}\) is at most \(\bigl(\ell_{\mathrm{sc}}+2D/r_{\mathrm{sc}}\bigr)|x-y|\).

We then Lipschitzize: let \(\ell_{\mathrm{lip}}\coloneqq 2\bigl(\ell_{\mathrm{sc}}+2D/r_{\mathrm{sc}}\bigr)\) and define
\[L_{\mathrm{lip}}(s)\coloneqq\inf_{u\le s}\bigl[L_{\mathrm{sc}}(u)+\ell_{\mathrm{lip}}(s-u)\bigr],\qquad
U_{\mathrm{lip}}(s)\coloneqq\sup_{u\ge s}\bigl[U_{\mathrm{sc}}(u)+\ell_{\mathrm{lip}}(s-u)\bigr].\]
Then \(L_{\mathrm{lip}}\le L_{\mathrm{sc}}\), \(U_{\mathrm{lip}}\ge U_{\mathrm{sc}}\), and both are \(\ell_{\mathrm{lip}}\)-Lipschitz nondecreasing.

We first bound the rightward shift of the level set of \(L_{\mathrm{lip}}\). Fix a constant \(c\) and take \(s\) with \(L_{\mathrm{lip}}(s)<c\). There exists \(u\le s\) such that \(L_{\mathrm{sc}}(u)+\ell_{\mathrm{lip}}(s-u)<c.\) If \(s-u\ge r_{\mathrm{sc}}\), then by the large-scale Lipschitzness of \(L_{\mathrm{sc}}\), \(c>L_{\mathrm{sc}}(u)+\ell_{\mathrm{lip}}(s-u)\ge L_{\mathrm{sc}}(s)+(\ell_{\mathrm{lip}}-\ell_{\mathrm{sc}})(s-u)\ge L_{\mathrm{sc}}(s),\) so \(s\in\{L_{\mathrm{sc}}<c\}\). If \(s-u<r_{\mathrm{sc}}\), then \(L_{\mathrm{sc}}(u)<c\), so \(u\in\{L_{\mathrm{sc}}<c\}\) and \(s<u+r_{\mathrm{sc}}\). In both cases \(s\le\sup\{L_{\mathrm{sc}}<c\}+r_{\mathrm{sc}}\), and therefore
\begin{equation}\label{eq:lip_shift}
\sup\{s:L_{\mathrm{lip}}(s)<c\}\le\sup\{s:L_{\mathrm{sc}}(s)<c\}+r_{\mathrm{sc}}.
\end{equation}

We now bound the threshold shift of \(U_{\mathrm{lip}}\). For every \(u\ge s_0\), we claim \(U_{\mathrm{sc}}(u)+\ell_{\mathrm{lip}}(s_0-u)\le U_{\mathrm{sc}}(s_0+r_{\mathrm{sc}}).\) If \(u-s_0\ge r_{\mathrm{sc}}\), then by the large-scale Lipschitzness of \(U_{\mathrm{sc}}\) with constant \(\ell_{\mathrm{lip}}/2=\ell_{\mathrm{sc}}+2D/r_{\mathrm{sc}}\), \(U_{\mathrm{sc}}(u)-U_{\mathrm{sc}}(s_0)\le\tfrac{\ell_{\mathrm{lip}}}{2}(u-s_0),\) so \(U_{\mathrm{sc}}(u)+\ell_{\mathrm{lip}}(s_0-u)\le U_{\mathrm{sc}}(s_0)-\tfrac{\ell_{\mathrm{lip}}}{2}(u-s_0)\le U_{\mathrm{sc}}(s_0+r_{\mathrm{sc}}).\) If \(u-s_0<r_{\mathrm{sc}}\), then \(U_{\mathrm{sc}}(u)\le U_{\mathrm{sc}}(s_0+r_{\mathrm{sc}})\) and \(\ell_{\mathrm{lip}}(s_0-u)\le0\), so the same inequality holds. Taking the supremum over \(u\ge s_0\) yields \(U_{\mathrm{lip}}(s_0)\le U_{\mathrm{sc}}(s_0+r_{\mathrm{sc}}).\) Combining this with \eqref{eq:lip_shift} and the monotonicity of \(\Gamma^{L_{\mathrm{sc}},U_{\mathrm{sc}}}\) in its first argument,
\[\Gamma^{L_{\mathrm{lip}},U_{\mathrm{lip}}}(s_0,t)\le\Gamma^{L_{\mathrm{sc}},U_{\mathrm{sc}}}(s_0+r_{\mathrm{sc}},t)+r_{\mathrm{sc}}\le s_0+T_\beta+2r_{\mathrm{sc}}-t\eta_\beta.\]

Finally, we mollify: take a standard mollifier \(\varphi_\sigma:[0,\sigma]\to\Rnn\) satisfying
\[
\varphi_\sigma\ge0,\qquad\int_0^\sigma\varphi_\sigma(u)\,du=1,\qquad\varphi_\sigma\in C^\infty,
\]
and define
\[L_\sigma(s)\coloneqq\int_0^\sigma\varphi_\sigma(u)L_{\mathrm{lip}}(s-u)\,du,\qquad
U_\sigma(s)\coloneqq\int_0^\sigma\varphi_\sigma(u)U_{\mathrm{lip}}(s+u)\,du.\]
Then \(L_\sigma\le L_{\mathrm{lip}}\), \(U_\sigma\ge U_{\mathrm{lip}}\), and both are smooth and nondecreasing.

Since \(L_\sigma(s)\) is a weighted average of \(L_{\mathrm{lip}}\) over \([s-\sigma,s]\), if \(L_\sigma(s)<c\), then there exists \(u\in[0,\sigma]\) with \(L_{\mathrm{lip}}(s-u)<c\), so \(s-u<\sup\{s':L_{\mathrm{lip}}(s')<c\}\) and hence
\begin{equation}\label{eq:moll_shift}
\sup\{s:L_\sigma(s)<c\}\le\sup\{s:L_{\mathrm{lip}}(s)<c\}+\sigma.
\end{equation}
Since \(U_{\mathrm{lip}}\) is nondecreasing, \(U_\sigma(s_0)\le U_{\mathrm{lip}}(s_0+\sigma)\). Combining \eqref{eq:moll_shift} with the monotonicity of \(\Gamma^{L_{\mathrm{lip}},U_{\mathrm{lip}}}\) in its first argument,
\[\Gamma^{L_\sigma,U_\sigma}(s_0,t)\le\Gamma^{L_{\mathrm{lip}},U_{\mathrm{lip}}}(s_0+\sigma,t)+\sigma
\le s_0+T_\beta+2r_{\mathrm{sc}}+2\sigma-t\eta_\beta.\]

Set \(L\coloneqq L_\sigma\) and \(U\coloneqq U_\sigma\). Then \(L\le\chi_1\le\chi_2\le U\), and \(L,U\) are smooth, nondecreasing, and Lipschitz. Moreover, \(\Gamma^{L,U}_\beta(s_0,t)\le s_0+T'_\beta-t\eta_\beta,\) where \(T'_\beta\coloneqq T_\beta+2r_{\mathrm{sc}}+2\sigma.\) Thus the envelope \((L,U)\) satisfies property \eqref{eq:propertyP} with the same decay rate \(\eta_\beta\) and an enlarged constant \(T'_\beta\). By the equivalence established in Section~\ref{sec:discrete}, the envelope automatically inherits the three geometric conditions, and its discriminative power is identical to that of the original comparison function pair.

\begin{remark}[Continuous-time case]\label{rem:envelope_continuous}
If the pair satisfies Condition~\ref{cond:g_continuous}, the envelopes inherit it: in logarithmic form the condition reads \(\chi_2(s)\le\chi_1(s+d)\) for all \(d>d_{\max}\), whence \(\chi_1(w)\ge\sup_{v<w-d_{\max}}\chi_2(v)\) and \(L_{\mathrm{mon}}(s+d)=\inf_{w\ge s+d}\chi_1(w)\ge\sup_{v\le s}\chi_2(v)=U_{\mathrm{mon}}(s)\) for \(d>d_{\max}\). The semi-continuous stage preserves this bound: for \(w\uparrow s+d\) and \(v\downarrow s\), eventually \(w-v>d_{\max}\), so \(L_{\mathrm{sc}}(s+d)\ge U_{\mathrm{sc}}(s)\). For Lipschitzization, \(L_{\mathrm{lip}}(v)\ge L_{\mathrm{sc}}(v-r_{\mathrm{sc}})\): each term \(L_{\mathrm{sc}}(u)+\ell_{\mathrm{lip}}(v-u)\) with \(u\le v\) is at least \(L_{\mathrm{sc}}(v-r_{\mathrm{sc}})\)---via the large-scale Lipschitz bound with \(\ell_{\mathrm{lip}}=2(\ell_{\mathrm{sc}}+2D/r_{\mathrm{sc}})\) when \(v-u\ge r_{\mathrm{sc}}\), and by monotonicity when \(v-u<r_{\mathrm{sc}}\); dually, \(U_{\mathrm{lip}}(s)\le U_{\mathrm{sc}}(s+r_{\mathrm{sc}})\). Mollification shifts by \(\sigma\) via convolution, so the final pair satisfies Condition~\ref{cond:g_continuous} with \(d_{\max}\) enlarged to \(d_{\max}+2r_{\mathrm{sc}}+2\sigma\).
\end{remark}

\begin{remark}[Every admissible continuous pair is one monotonization away from the Panteley--Lor\'ia domain]\label{rem:pl_reachable}
Apply the monotonization step to the lower bound only: \(L_{\mathrm{mon}}(s)\coloneqq\inf_{u\ge s}\chi_1(u)\) is nondecreasing, \(L_{\mathrm{mon}}\le\chi_1\le\chi_2\), and by \eqref{eq:mono_invariance} the pair \((L_{\mathrm{mon}},\chi_2)\) satisfies property \eqref{eq:propertyP} exactly when \((\chi_1,\chi_2)\) does. Hence, for every pair admissible for the continuous-time criterion, the lower-monotonized pair---nondecreasing below, arbitrary above---inherits property \eqref{eq:propertyP}, satisfies Conditions~\ref{cond:package} and \ref{cond:g_continuous} by Theorem~\ref{thm:continuous}, and therefore satisfies the direct Panteley--Lor\'ia conditions by the converse direction of Proposition~\ref{prop:pl_equiv} (which requires monotonicity of the lower function only). In this sense, the direct form of the Panteley--Lor\'ia criterion is reached from every admissible continuous pair by a single monotonization; cf. Remark~\ref{rem:pl_complete}.
\end{remark}

\begin{remark}[Strictly increasing envelopes]\label{rem:strict_envelope}
In the discrete-time case, the lower envelope can be further polished to a strictly increasing, continuous, and Lipschitz one---with \(U\) unchanged and Condition~\ref{cond:g_general} preserved---provided the left tail \(\chi_2(-\infty)=-\infty\) is in force and \(\chi_1(+\infty)=+\infty\) (which excludes exactly the bounded-envelope case of Section~\ref{subsec:tails}). In the continuous-time case this is impossible in general: a pair with plateaus of unbounded length is a counterexample. \ifextended The construction, its proof, and the counterexample are given in Appendix~\ref{app:strict}.\else The construction, its proof, and the counterexample are given in \cite[Appendix~C]{guo2026extended}.\fi
\end{remark}

\subsection{An integral-form check and verifiable classes}
\label{subsec:integral_tool}

Let \(L,U\) be the continuous, nondecreasing, and Lipschitz envelopes constructed in Section~\ref{subsec:envelope_construction}. Define
\[
\tau(s)\coloneqq\inf\left\{t\ge0:\int_s^{s+t}\bigl(L(u)-L(s)\bigr)du=1\right\}.
\]
If the set is empty, we set \(\tau(s)=+\infty\) and \(1/\tau(s)=0\).

\begin{theorem}[Integral-form tool]\label{thm:integral_discrete}
Let the underlying pair satisfy Conditions~\ref{cond:bounded_close}, \ref{cond:large_lip}, and \ref{cond:g_general}, and let \(L,U\) and \(\tau\) be the envelopes and the gap time defined above. There exists a continuous, strictly increasing function \(h:\R_{\ge0}\to\R_{\ge0}\) satisfying \(h(0)=0\) such that
\[
U(s)-L(s)\le h\!\left(\frac{1}{\tau(s)}\right),\quad\forall s\in\R.
\]
\end{theorem}

\paragraph{Proof of Theorem~\ref{thm:integral_discrete}}

In this proof and the next, \(g(d,s)\coloneqq L(s+d)-U(s)\) and \(G(d)\coloneqq\inf_s g(d,s)\) denote the displacement function and joint lower envelope of the envelope pair \((L,U)\), which inherits the relevant condition from Section~\ref{subsec:envelope_construction} (preservation of \eqref{eq:propertyP} and Theorem~\ref{thm:strict_beta}). Since \(L\le\chi_1\le\chi_2\le U\), we have \(U(s)-L(s)\ge0\). Suppose, for contradiction, that there exist \(\varepsilon>0\) and a sequence \(s_n\) such that \(\tau(s_n)\to\infty\) but \(U(s_n)-L(s_n)\ge\varepsilon\). If \(\tau(s_n)=+\infty\) for infinitely many \(n\), the \(\tau=+\infty\) paragraph below forces \(U(s_n)=L(s_n)\), contradicting \(U(s_n)-L(s_n)\ge\varepsilon\); passing to a subsequence, we may assume \(\tau_n\coloneqq\tau(s_n)<+\infty\). By the definition of \(\tau_n\), \(\int_{s_n}^{s_n+\tau_n}(L(u)-L(s_n))\,du=1.\) Take \(n\) sufficiently large so that \(\tau_n\ge4/\varepsilon\). If \(L(s_n+\tau_n/2)>L(s_n)+\varepsilon/2\), then by the monotonicity of \(L\), for every \(u\in[s_n+\tau_n/2,s_n+\tau_n]\) we have \(L(u)-L(s_n)>\varepsilon/2\), so the integral over the right half exceeds \(\tau_n\varepsilon/4\ge1\), contradicting the fact that the total integral equals 1. Hence \(L(s_n+\tau_n/2)\le L(s_n)+\varepsilon/2.\) Since \(U(s_n)\ge L(s_n)+\varepsilon\), we obtain \(g(\tau_n/2,s_n)=L(s_n+\tau_n/2)-U(s_n)\le-\varepsilon/2<0.\) Since \(\tau_n/2\to\infty\), this contradicts the condition \(G_{\lim}\ge0\) in Condition~\ref{cond:g_general}. The case \(\tau(s)=+\infty\) is similar: the integral then stays below \(1\) for every \(t\); since it is at least \(\tfrac{t}{2}\bigl(L(s+\tfrac{t}{2})-L(s)\bigr)\), \(L(s+d)-L(s)<4/d\to0\); in fact, for each fixed \(d>0\) and any \(D>d\), monotonicity gives \(0\le L(s+d)-L(s)\le L(s+D)-L(s)<4/D\), so \(L(s+d)=L(s)\): \(L\) is constant on \([s,+\infty)\), and \(g(d,s)=L(s+d)-U(s)\to L(s)-U(s)\), so \(G_{\lim}\ge0\) forces \(U(s)=L(s)\). The contradiction argument above shows that for every \(\varepsilon>0\) there exists \(T_\varepsilon\) such that \(\tau(s)>T_\varepsilon\) implies \(U(s)-L(s)<\varepsilon\); equivalently, \(U(s)-L(s)\to0\) uniformly as \(\tau(s)\to\infty\). Define \(w(s)=U(s)-L(s)\) and \(q(s)=1/\tau(s)\). Then \(w(s)\to0\) uniformly as \(q(s)\to0^+\). Hence there exists a continuous, strictly increasing function \(h\) satisfying \(h(0)=0\) such that \(w(s)\le h(q(s))\): \(w\) is globally bounded by the envelope's closeness constant, so \(h_1(q)\coloneqq\sup\{w(s):q(s)\le q\}\) is finite and nondecreasing, and \(h_1(0^+)=0\) since \(w(s)\to0\) uniformly as \(q(s)\to0^+\); setting \(h_2(x)\coloneqq h_1(x)+x\) and \(h_3(x)\coloneqq x+\frac{1}{x}\int_x^{2x}h_2(y)\,dy\) for \(x>0\) gives a continuous, strictly increasing majorant \(h_3\ge h_2\ge h_1\) (strictly increasing as the sum of \(x\) and a nondecreasing average); since the integral average is at most \(h_2(2x)\to0\) as \(x\to0^+\), the extension \(h_3(0)=0\) is continuous.

\begin{theorem}[Integral-form tool, continuous-time]\label{thm:integral_continuous}
Let the pair satisfy Condition~\ref{cond:g_continuous}, so the envelopes inherit it (Remark~\ref{rem:envelope_continuous}). Then the conclusion of Theorem~\ref{thm:integral_discrete} holds as stated, with \(1/\tau(s)=0\) when \(\tau(s)=+\infty\), and in this case \(U(s)-L(s)=0\) is required.
\end{theorem}

\paragraph{Proof of Theorem~\ref{thm:integral_continuous}}

By Condition~\ref{cond:g_continuous}, there exists \(d_{\max}\ge0\) such that \(G(d)\ge0\) for \(d>d_{\max}\). If \(\tau(s)=+\infty\), then for every \(t\ge0\), \(\int_s^{s+t}(L(u)-L(s))\,du<1.\) If \(U(s)-L(s)>0\), we claim that \(L(s+d)-U(s)<0\) for all sufficiently large \(d\). Indeed, \(\tau(s)=+\infty\) gives \(\int_s^{s+t}(L(u)-L(s))\,du<1\) for every \(t\ge0\); since \(L\) is nondecreasing, the integral is at least \(\frac{t}{2}\bigl(L(s+t/2)-L(s)\bigr)\), hence \(L(s+t/2)-L(s)<2/t\), i.e., \(L(s+d)-L(s)<4/d\to0\); for each fixed \(d>0\) and any \(D>d\), monotonicity gives \(0\le L(s+d)-L(s)\le L(s+D)-L(s)<4/D\), so \(L(s+d)=L(s)\): \(L\) is constant on \([s,+\infty)\). Therefore \(L(s+d)-U(s)\to L(s)-U(s)<0\), so \(g(d,s)<0\) for all sufficiently large \(d\), which contradicts \(G(d)\ge0\) for \(d>d_{\max}\). Hence in this case \(U(s)-L(s)=0\). For the case \(\tau(s)<+\infty\), the proof is the same as that of Theorem~\ref{thm:integral_discrete}.

\begin{remark}
If \(U(s)\to+\infty\), then \(L(s)\to+\infty\) as well (the gap being bounded); the integral exceeds \(1\) for large \(t\), and \(\tau\) is always finite, so the convention \(1/\tau=0\) is never active.
\end{remark}

Integral characterizations of uniform asymptotic and exponential stability at the system level were given by Teel, Panteley and Lor\'ia \cite{teel2002}. Our integral test is at the criterion level: the gap time \(\tau(s)\) involves only the pair \((\chi_1,\chi_2)\), not system trajectories---a necessary check, usable as a single integral test rejecting a candidate pair before the full three-condition check (it does not certify one)---and a cheap one, \(\tau(s)\) being one-dimensional whereas verifying \(G_{\lim}\ge0\) or \(G(d)\ge0\) directly requires a two-dimensional global infimum. In practice: if \(\tau(s_n)\to\infty\) while \(U(s_n)-L(s_n)\) does not tend to \(0\), the pair is rejected.

\paragraph{Verifiable classes}
\label{subsec:verifiable}

The three conditions apply to any analytically derived bounding pair---the tight radial envelopes of \eqref{eq:radial} need not be computed---tightening the pair enlarges \(g\) pointwise and shrinks the gap \(\chi_2-\chi_1\), so the tight envelopes offer the best chance of verification, and the failure of a looser pair indicts the pair, not the Lyapunov function---and involve global infima over \(s\) and \(d\), so one may ask whether the verification merely relocates an undecidable problem. We do not claim general decidability; on two common function classes, however, the verification is an explicit computation. \ifextended The plateau pair of Appendix~\ref{app:strict}\else The plateau pair of \cite[Appendix~C]{guo2026extended}\fi\ (whose \(G\equiv0\)) falls into the first class.

\paragraph{Piecewise-power pairs} Suppose that \(\chi_1,\chi_2\) are continuous and piecewise linear with a breakpoint set \(\{a_i\}\) (a generic notation, unrelated to the specific points of the plateau pair \ifextended of Appendix~\ref{app:strict}\else of \cite{guo2026extended}\fi) having no finite accumulation point (equivalently, \(\alpha_1,\alpha_2\) are piecewise power functions). For fixed \(d\ge0\), the function \(s\mapsto g(d,s)=\chi_1(s+d)-\chi_2(s)\) is again piecewise linear, with breakpoints contained in \(\{a_i\}\cup\{a_i-d\}\), so
\[G(d)=\inf_s g(d,s)=\min\bigl\{\inf_i g(d,a_i),\ \inf_i g(d,a_i-d),\ \underline g_{+\infty}(d),\ \underline g_{-\infty}(d)\bigr\},\]
where \(\underline g_{\pm\infty}(d)\coloneqq\liminf_{s\to\pm\infty}g(d,s)\) are read off the piecewise linear form directly; if the terminal slopes \(p_1,p_2\) of \(\chi_1,\chi_2\) at \(+\infty\) exist, then \(\underline g_{+\infty}(d)=+\infty\) if \(p_1>p_2\), \(\underline g_{+\infty}(d)=-\infty\) if \(p_1<p_2\), and if \(p_1=p_2\) the tail of \(g(d,\cdot)\) is bounded and \(\underline g_{+\infty}(d)\) is its finite liminf; the \(-\infty\) end is symmetric, with the inequality reversed. Hence \(G(d)\), \(G_{\lim}\), and the boundedness of \(\{d:G(d)<0\}\) all reduce to countable explicit evaluations together with a sign test on the terminal slopes. The variable-power pair of Remark~\ref{rem:varpower} lies in this class. If, in addition, the slopes are asymptotically periodic, the candidate breakpoint set is finite over one period, and the verification reduces to finitely many evaluations.

\paragraph{Polynomial pairs} For polynomial or rational comparison functions, semi-algebraic relaxations of the three conditions can be checked by sum-of-squares programming, complementing the system-level implementation of \cite{jagt2026}; a criterion-level formulation is left as future work.

\subsection{A worked example: envelopes and the integral check}
\label{subsec:ex_sawtooth}

Take the pair
\[
\chi_1(s)=s+2\sin s,\qquad \chi_2(s)=\chi_1(s)+1,
\]
with tails \(\chi_1(\pm\infty)=\pm\infty\). Bounded closeness holds with \(D=1\), and \(\chi_1\) is globally \(3\)-Lipschitz. Moreover,
\[
g(d,s)=d+2\sin(s+d)-2\sin s-1\ge d-5,
\]
so \(G(d)\ge0\) for all \(d\ge d_{\max}\coloneqq5\), and Condition~\ref{cond:g_continuous} holds: by Theorem~\ref{thm:continuous} the continuous-time general criterion covers this pair. The Panteley--Lor\'ia criterion as stated in \cite{panteley2010} is not applicable, since \(\alpha_1(r)=r\,e^{2\sin\ln r}\) is not monotone and the inverse does not exist; the direct conditions \eqref{eq:pl_direct} themselves do hold (e.g., with \(c_1=e^6\), \(c_2=e^{-4}\), \(c_3=e^{-1}\): \(\ln c_1=6>d_{\max}=5\) gives the first, and \(\gamma=D+\Lambda r_0=4\), \(c=r_0=1\) the second).

The envelope construction of Section~\ref{subsec:envelope_construction} can be followed in closed form. The local minima of \(\chi_1\) are at \(m_k=\tfrac{4\pi}{3}+2k\pi\) with values \(m_k-\sqrt3\), and the local maxima at \(M_k=\tfrac{2\pi}{3}+2k\pi\) with values \(M_k+\sqrt3\). Monotonization gives floor-and-ramp envelopes: \(L_{\mathrm{mon}}\) follows \(\chi_1\) up to a breakpoint \(s^*_k\) on each increasing ramp, then stays flat at \(\chi_1(m_{k+1})\) until the next minimum, with \(\chi_1(s^*_k)=\chi_1(m_{k+1})\) defining \(s^*_k\) implicitly; dually, \(U_{\mathrm{mon}}\) follows \(\chi_2\) up to each local maximum \(M_k\), then stays flat at \(\chi_2(M_k)\) until the point \(t^*_k\) on the next ramp with \(\chi_2(t^*_k)=\chi_2(M_k)\). Lipschitzization removes the kinks and mollification smooths the result, the level-set shifts bounded by \eqref{eq:lip_shift} and \eqref{eq:moll_shift}.

On this envelope the integral check of Theorem~\ref{thm:integral_continuous} is verified directly: for \(s\) at the start of a flat, \(\tau(s)\) is the flat's length plus the recovery time from the quadratic restart at the next minimum (where \(\chi_1''=\sqrt3\), so the unit integral needs \((\sqrt3/6)T^3=1\)), namely \((2\sqrt3)^{1/3}\); the flats have bounded length (\(\le2\pi\)), so \(\tau\) is bounded above and \(1/\tau\) is bounded below, while \(U_{\mathrm{mon}}-L_{\mathrm{mon}}\) is uniformly bounded (by the envelope's closeness constant), say by \(C\); with \(\tau\le\bar T\), the linear function \(h(x)=C\bar T x\) then witnesses the theorem's inequality. The pair lies outside the dynamics part of Remark~\ref{rem:realizability} (whose semiflow requires a strictly increasing pair), but it is a legitimate object of the theory at the level of pairs, where the criterion operates.

\begin{remark}[Sharpness of the funnel]\label{rem:sharpness}
The two-sided level-set bound behind every conclusion is sharp already for constant-ratio pairs. For \(\alpha_1(r)=r^2\), \(\alpha_2(r)=3r^2\) one has \(\{r<\sqrt{c/3}\}\subseteq\{V<c\}\subseteq\{r<\sqrt{c}\}\) at \(t=0\), and both inclusions are equalities for admissible Lyapunov functions (\(V=3r^2\) and \(V=r^2\), respectively): with only the pair in hand, neither endpoint of the funnel can be moved inward.
\end{remark}

\section{Conclusions}
\label{sec:conclusion}

We have established necessary and sufficient structural conditions for the Lyapunov-type exponential stability criterion, making it both admissible and tight. In logarithmic coordinates, the theory reduces to a single two-sided displacement function of the pair: in discrete time, migration holds if and only if the regularity package and the joint lower bound hold simultaneously; continuous time strengthens the latter, and the two are separated exactly by the minimum step size. Written without its inverse-function formulation, the Panteley--Lor\'ia criterion coincides with our continuous-time conditions on the nondecreasing pairs and loses soundness beyond them: it was already complete on the classical regularity class. The envelope construction shows that regularization costs no discriminative power, retroactively legitimizing the smoothness, monotonicity, and Lipschitz conventions imposed on comparison functions since Hahn; radial unboundedness is likewise dispensable. The criterion level is the unique level at which the characterization is both exact and dimension-free. Future work includes nonautonomous \cite{karafyllis2003}, switched, time-delay, and stochastic extensions, non-exponential rates \cite{appleby2016}, and numerical verification.

\ifextended
\appendix

This extended version collects the material omitted from the journal version for length, in the order in which it is cited there: the variable-power pair of Remark~\ref{rem:varpower} (Appendix~\ref{app:varpower}), the closed-form analysis of the skeleton pair of Remark~\ref{rem:trench_rate} (Appendix~\ref{app:skeleton}), and the strictly increasing lower envelope of Remark~\ref{rem:strict_envelope} together with the continuous-time counterexample (Appendix~\ref{app:strict}). Appendix~\ref{app:numerical} contains numerical examples and Appendix~\ref{app:beta} a remark on the strict-inequality convention; these two are not cited in the journal version.

\section{The variable-power pair of Remark~\ref{rem:varpower}}
\label{app:varpower}

Remark~\ref{rem:varpower} asserts that the three conditions require neither fixed powers nor fixed slopes, while Hahn's same-power bounds fail for the pair exhibited. This appendix constructs the pair and proves the assertions.

Define $\alpha:\Rnn\to\Rnn$ in the radial coordinate $r=\norm{x}$, with $\alpha(0)=0$ and, for $j=0,1,2,\dots$,
\begin{equation}\label{eq:piecewise}
\alpha(r)=
\begin{cases}
r, & r\in[0,2)\cup\bigcup_{j=0}^{\infty}\left[2^{6\cdot8^j},\,2^{8^{j+1}}\right),\\[6pt]
r^2, & r\in\bigcup_{j=0}^{\infty}\left[2^{\frac32\cdot8^j},\,2^{2\cdot8^j}\right),\\[6pt]
2^{-3\cdot8^j}\,r^4, & r\in\bigcup_{j=0}^{\infty}\left[2^{8^j},\,2^{\frac32\cdot8^j}\right),\\[6pt]
2^{3\cdot8^j}\,r^{1/2}, & r\in\bigcup_{j=0}^{\infty}\left[2^{2\cdot8^j},\,2^{6\cdot8^j}\right).
\end{cases}
\end{equation}
The four interval families are mutually disjoint and cover $\Rnn$. At each joint the values from the two sides agree (e.g., $2^{-3\cdot8^j}\bigl(2^{8^j}\bigr)^4=2^{8^j}$ and $2^{3\cdot8^j}\bigl(2^{6\cdot8^j}\bigr)^{1/2}=2^{6\cdot8^j}$), so $\alpha$ is continuous; each piece being strictly increasing, $\alpha\in\mathcal{K}_\infty$. The local powers cycle through $1,2,4,\tfrac12$ over intervals of exponentially growing lengths, so the power is not fixed.

\begin{proposition}\label{prop:varpower}
Let $\alpha_1=\alpha_2=\alpha$ with $\alpha$ of \eqref{eq:piecewise}. Then:
\begin{enumerate}
\item the pair satisfies Conditions~\ref{cond:bounded_close}, \ref{cond:large_lip}, and \ref{cond:g_general}; indeed the strengthened Condition~\ref{cond:g_continuous} holds with $d_{\max}=0$;
\item no fixed power $p>0$ and constants $\eta_1,\eta_2>0$ with $\eta_1 r^p\le\alpha(r)\le\eta_2 r^p$ exist;
\item the criterion's conclusion is verified directly: if along a trajectory $V(r_{k+1})\le e^{-\beta}V(r_k)$, then $r_{k+1}\le e^{-\beta/4}r_k$, and the rate is sharp.
\end{enumerate}
\end{proposition}

\begin{proof}
Since $\alpha_1=\alpha_2=\alpha$, the sandwich $\alpha(\norm{x})\le V(x)\le\alpha(\norm{x})$ pins $V(x)=\alpha(\norm{x})$. In logarithmic coordinates, $\chi(s)=\ln\alpha(e^s)$ is continuous and piecewise linear, with slopes in $\{1,2,4,\tfrac12\}$ on the intervals of \eqref{eq:piecewise}.

(i) $\chi_2-\chi_1=0$, so Condition~\ref{cond:bounded_close} holds with any $D>0$. Every slope of $\chi$ lies in $[\tfrac12,4]$, and $\chi$ is continuous and piecewise linear, hence globally $4$-Lipschitz: Condition~\ref{cond:large_lip} holds. Moreover, every slope being at least $\tfrac12$, for $d\ge0$,
\[
g(d,s)=\chi(s+d)-\chi(s)\ge\frac{d}{2}\ge0,
\]
so $G(d)\ge0$ identically: $G_{\min}\ge0$, $G_{\lim}\ge0$, and Condition~\ref{cond:g_continuous} holds with $d_{\max}=0$.

(ii) For $p<1$, $\alpha(r)/r^p=r^{1-p}\to\infty$ along the slope-$1$ rays, so no upper bound $\eta_2r^p$ exists. For $p>1$, $\alpha(r)/r^p=r^{1-p}\to0$ along the same rays, so no positive lower bound $\eta_1r^p$ exists. For $p=1$, at $r_j=2^{4\cdot8^j}$ (inside the slope-$\tfrac12$ interval $\left[2^{2\cdot8^j},2^{6\cdot8^j}\right)$), $\alpha(r_j)/r_j=2^{3\cdot8^j}r_j^{-1/2}=2^{8^j}\to\infty$.

(iii) The premise $V(r_{k+1})\le e^{-\beta}V(r_k)$ reads $\chi(s_{k+1})\le\chi(s_k)-\beta$ in logarithmic coordinates, with $s_k=\ln r_k$. Since $\chi$ is increasing, $s_{k+1}<s_k$, and since $\chi$ is globally $4$-Lipschitz,
\[
\beta\le\chi(s_k)-\chi(s_{k+1})\le4\,(s_k-s_{k+1}),
\]
so $s_{k+1}\le s_k-\beta/4$, i.e., $r_{k+1}\le e^{-\beta/4}r_k$. The rate is sharp: just inside the next power-$4$ interval, at $r_k=2^{8^{j+1}}e^{\beta/4}$, a direct evaluation shows the ratio $r_{k+1}/r_k$ attains $e^{-\beta/4}$.
\end{proof}

\section{The skeleton pair of Remark~\ref{rem:trench_rate}}
\label{app:skeleton}

Remark~\ref{rem:trench_rate} asserts that the pair exhibited there satisfies $G(d)<0$ at every $d>0$, $G_{\lim}=0$, and $-G(d)\sim 2A(\ln d)^{-2}$ as $d\to\infty$. This appendix proves the assertions. Recall the skeleton $\chi_0$ of Remark~\ref{rem:trench_rate}, restated here for convenience:
\begin{equation}\label{eq:skeleton}
\chi_0(s)=
\begin{cases}
s, & s\le a_0,\\[3pt]
\eta_j-e^{-j^2}(s-a_j), & s\in[a_j,b_j],\\[3pt]
\eta_j-e^{-j^2}(b_j-a_j)+2(s-b_j), & s\in[b_j,a_{j+1}],
\end{cases}
\end{equation}
where $a_j=2e^j$, $b_j=a_{j+1}-1+\varepsilon\sin j$ with $\varepsilon=0.3$, and $\eta_0=a_0$, $\eta_{j+1}=\eta_j-e^{-j^2}(b_j-a_j)+2(a_{j+1}-b_j)$. The decreasing segments have lengths $D_j\coloneqq b_j-a_j=2e^j(e-1)-1+\varepsilon\sin j\sim2(e-1)e^j$ and slopes $-e^{-j^2}$, while the increasing segments have slope $2$ and length $1-\varepsilon\sin j\in[0.7,1.3]$. Recall also the amplitude
\[
\bar u(s)=
\begin{cases}
A, & s\le a_0,\\[3pt]
(1-\lambda)\dfrac{A}{(j+1)^2}+\lambda\dfrac{A}{(j+2)^2}, & s\in(a_j,a_{j+1}],\ \lambda=\dfrac{s-a_j}{a_{j+1}-a_j},
\end{cases}
\]
with $A=0.15$, and the pair $\chi_1=\chi_0-\bar u$, $\chi_2=\chi_0+\bar u$. Finally, recall the monotone envelopes
\[
L(v)\coloneqq\inf_{t\ge v}\chi_1(t),\qquad U(s)\coloneqq\sup_{t\le s}\chi_2(t),
\]
and write $g(d,s)\coloneqq L(s+d)-U(s)$, $G(d)\coloneqq\inf_s g(d,s)$. Then $L,U$ are continuous and nondecreasing, $L\le\chi_1\le\chi_2\le U$, and both tails are attained.

\begin{proposition}\label{prop:skeleton}
The pair $(L,U)$ satisfies {\rm(i)} $G(d)<0$ for every $d>0$; {\rm(ii)} $\inf_{d\ge0}G(d)>-\infty$; and {\rm(iii)} $-G(d)\sim2A(\ln d)^{-2}$ as $d\to\infty$. Hence $(L,U)$ satisfies the discrete-time Condition~\ref{cond:g_general} with $G_{\lim}=0$, while $\{d\ge0:G(d)<0\}=(0,\infty)$ is unbounded, so the continuous-time Condition~\ref{cond:g_continuous} fails.
\end{proposition}

\begin{proof}
Write $E_j\coloneqq e^{-j^2}D_j$ (the drop of $\chi_0$ on the $j$-th decreasing segment) and $\Delta_j\coloneqq\sum_{k\ge j}E_k$; since $E_j\sim2(e-1)e^{j-j^2}$ decays double-exponentially, $\Delta_0<\infty$ and $\Delta_j=o(j^{-2})$. Let $j(s)$ denote the cycle containing $s$, i.e., $s\in(a_{j(s)},a_{j(s)+1}]$. Three elementary facts follow from the definitions. First,
\begin{equation}\label{eq:sk_f1}
\chi_0(t)\ge\chi_0(s)-\Delta_{j(s)}\qquad(t\ge s),
\end{equation}
since between $s$ and $t$ the function $\chi_0$ drops only on decreasing segments. Second, on each cycle,
\begin{equation}\label{eq:sk_f2}
\frac{A}{(j+2)^2}\le\bar u(s)\le\frac{A}{(j+1)^2},\qquad s\in(a_j,a_{j+1}].
\end{equation}
Third, the increments satisfy
\[
\eta_{j+1}-\eta_j=2(1-\varepsilon\sin j)-E_j\ge1.4-E_j>0\quad(j\ge2),
\]
so $(\eta_j)$ is increasing for $j\ge2$; a direct evaluation gives $\eta_0=2$, $\eta_1\approx1.56$, $\eta_2\approx-0.10$, $\eta_3\approx0.90$, $\eta_4\approx2.81$, so $\eta_j\ge\eta_0$ for $j\ge4$, and $\eta_j\to+\infty$. Consequently, for $j(s)\ge4$,
\begin{equation}\label{eq:sk_f3}
U(s)\le\chi_0(s)+E_{j(s)}+\frac{A}{j(s)^2}:
\end{equation}
for $t$ on the cycle of $s$ with $t\le s$, the maximum of $\chi_0$ on that part is at most $\eta_{j(s)}$, so $\chi_2(t)\le\eta_{j(s)}+A/(j(s)+1)^2\le\chi_0(s)+E_{j(s)}+A/(j(s)+1)^2$, since $\chi_0(s)\ge\eta_{j(s)}-E_{j(s)}$; for $t$ on an earlier cycle $k<j(s)$, if $k\ge2$ then $\chi_2(t)\le\eta_{k+1}+A/(k+1)^2\le\eta_{j(s)}+A/j(s)^2$, because $\eta_{m+1}-\eta_m\ge0.9$ for $m\ge2$ while $A/(k+1)^2-A/j(s)^2\le A/9<0.02$, and if $k\le1$ then $\chi_2(t)\le\eta_0+A=2.15<\eta_4-E_4\le\chi_0(s)$, since $\eta_4\approx2.81$.

\emph{Upper bound, proving \textup{(i)} and the upper half of \textup{(iii)}.} For $d\ge0$ let $j(d)\coloneqq\min\{j\ge1:D_j\ge d\}$, which exists since $D_j\to\infty$; then $b_{j(d)}-d\in[a_{j(d)},b_{j(d)}]$. Since $L\le\chi_1$ and $U\ge\chi_2$, and both points lie on the same decreasing segment,
\[
\begin{aligned}
G(d)&\le\chi_1(b_{j(d)})-\chi_2(b_{j(d)}-d)
=-e^{-j(d)^2}d-\bar u(b_{j(d)})-\bar u(b_{j(d)}-d)\\
&\le-\frac{2A}{(j(d)+2)^2}<0,
\end{aligned}
\]
using \eqref{eq:sk_f2}. Since $D_j\sim2(e-1)e^j$, $j(d)=\ln d-\ln(2(e-1))+o(1)$, so $(j(d)+2)^2=(\ln d)^2(1+o(1))$ and
\begin{equation}\label{eq:sk_upper}
-G(d)\ge\frac{2A(1-o(1))}{(\ln d)^2}\qquad(d\to+\infty).
\end{equation}

\emph{Lower bound, proving the lower half of \textup{(iii)}.} Fix $d\ge a_4$ and $s\in\R$, and write $j=j(s)$, $k=j(s+d)$; then $k\ge4$. From \eqref{eq:sk_f1}, \eqref{eq:sk_f2},
\begin{equation}\label{eq:sk_L}
L(v)\ge\chi_0(v)-\Delta_{j(v)}-\frac{A}{(j(v)+1)^2}.
\end{equation}

Suppose first that $j\le3$, i.e., $s\le a_4$; this case requires a uniform estimate since $s$ may drift to $-\infty$ as $d$ grows. If $s\le a_0$, then $\chi_2(t)=t+A$ for $t\le s$, so $U(s)=s+A$ exactly, while \eqref{eq:sk_L} at $v=s+d$ gives $L(s+d)\ge\chi_0(s+d)-\Delta_0-A$; hence, writing $u\coloneqq s+d$,
\[
g(d,s)\ge\chi_0(u)-u+d-\Delta_0-2A.
\]
On each decreasing segment $\chi_0(u)-u$ has slope $-1-e^{-j^2}<0$, and on each increasing segment slope $+1$, so its local minima occur at the points $b_j$, with values $\eta_j-E_j-b_j$; since $\eta_j$ grows linearly while $b_j$ grows exponentially, the minimum of $\chi_0(u)-u$ over $u\le a_0+d$ equals $-(a_0+d)+O(\ln d)$, and therefore
\[
g(d,s)\ge\ln d\qquad(s\le a_0,\ d\ \text{large}).
\]
If $a_0\le s\le a_4$, then $U(s)\le\eta_4+A$ and $s+d\ge a_0+d$, so $j(s+d)\to\infty$ as $d\to\infty$ and \eqref{eq:sk_L} gives $L(s+d)\ge\eta_{j(s+d)}-E_{j(s+d)}-\Delta_{j(s+d)}-A/(j(s+d)+1)^2\to+\infty$; hence $g(d,s)\ge\ln d$ as well for large $d$, uniformly on $[a_0,a_4]$. In both subcases $g(d,s)$ is eventually positive, so the range $j(s)\le3$ contributes nothing to the asymptotics of $G(d)$. Assume now $j\ge4$.

If $k\ge j+2$, then $\chi_0(s+d)\ge\eta_k-E_k$ and $\chi_0(s)\le\eta_{j+1}$, so by \eqref{eq:sk_L} and \eqref{eq:sk_f3},
\[
g(d,s)\ge(\eta_{j+2}-\eta_{j+1})-E_k-\Delta_k-\frac{A}{(k+1)^2}-E_j-\frac{A}{j^2}\ge\tfrac12
\]
for $j\ge4$, the increment $\eta_{j+2}-\eta_{j+1}$ being at least $0.9$ while all remaining terms are $o(1)$.

If $k=j+1$, then $\chi_0(s)\le\eta_{j+1}$ and $\chi_0(s+d)\ge\eta_{j+1}-E_{j+1}$, so
\[
g(d,s)\ge-E_{j+1}-\Delta_{j+1}-\frac{A}{(j+2)^2}-E_j-\frac{A}{j^2}
=-\frac{2A+o(1)}{j^2}.
\]

If $k=j$, which requires $d\le D_j$, i.e., $j\ge j(d)$, then $\chi_0(s+d)-\chi_0(s)\ge-E_j$, and the same bound $g(d,s)\ge-(2A+o(1))/j^2$ holds with $j\ge j(d)$. Combining all cases and using $j(d)\to\infty$ as $d\to\infty$,
\begin{equation}\label{eq:sk_lower}
G(d)\ge-\frac{2A(1+o(1))}{(\ln d)^2}\qquad(d\to+\infty).
\end{equation}
Together with \eqref{eq:sk_upper} this gives $-G(d)\sim2A(\ln d)^{-2}$.

\emph{Finiteness, proving \textup{(ii)}.} For every $d\ge0$ and $s\in\R$, \eqref{eq:sk_f1} gives $L(s+d)\ge\chi_0(s)-\Delta_0-A$, while the global bound $U(s)\le\chi_0(s)+E_0+(\eta_0-\eta_2)+A$ holds for all $s$ (the term $\eta_0-\eta_2$ covers the non-monotone initial segment of $(\eta_j)$). Hence
\[
g(d,s)\ge-2\Delta_0-E_0-(\eta_0-\eta_2)-2A>-\infty,
\]
so $\inf_{d\ge0}G(d)>-\infty$.
\end{proof}

\section{The strictly increasing envelope of Remark~\ref{rem:strict_envelope}}
\label{app:strict}

This appendix completes Remark~\ref{rem:strict_envelope}: we restate and prove the discrete-time strictly increasing envelope construction, and then give the closed-form counterexample showing that no such envelope exists in general in continuous time.

\begin{proposition}\label{prop:strict}
Let $(\chi_1,\chi_2)$ satisfy Conditions~\ref{cond:bounded_close}, \ref{cond:large_lip}, and \ref{cond:g_general}, with the left tail $\chi_2(-\infty)=-\infty$ in force and with $\chi_1(+\infty)=+\infty$ (so that $L$ has no right-unbounded plateau). Then there exists a strictly increasing lower envelope $L_{\mathrm{strict}}\le L$---only the lower envelope is replaced, $U$ being unchanged---which is continuous and Lipschitz, and Condition~\ref{cond:g_general} is preserved.
\end{proposition}

\begin{remark}[Bounded envelopes]\label{rem:bounded}
The hypothesis $\chi_1(+\infty)=+\infty$ of Proposition~\ref{prop:strict} excludes a right-unbounded plateau of $L$, which occurs exactly for bounded envelopes (Section~\ref{subsec:tails}). In that case a strictly increasing minorant exists trivially---$L$ minus a positive strictly decreasing function---but the preservation of Condition~\ref{cond:g_general} requires a separate analysis, which we do not pursue here.
\end{remark}

The construction is performed on the final envelope $L$ of Section~\ref{subsec:envelope_construction}, which is smooth, nondecreasing, and Lipschitz with constant $\ell_{\mathrm{lip}}$. Let the plateaus of $L$ be $[a_i,b_i]$ with value $p_i$. Since $L$ is nondecreasing, its plateaus are at most countable in number (each contains a rational number). Moreover, the plateau family is locally finite: $L$ is the fixed-width mollification of $L_{\mathrm{lip}}$ with kernel width $\sigma$, and since the mollifier is positive a.e.\ on its support, $L$ is constant on $[a,b]$ exactly when $L_{\mathrm{lip}}$ is constant on $[a-\sigma,b]$; the plateaus of $L$ thus arise from plateaus of $L_{\mathrm{lip}}$ of length exceeding $\sigma$, and consecutive plateaus of $L$ are separated by gaps exceeding $\sigma$ (a smaller gap would merge the two pre-images into a single plateau of $L$, contradicting maximality), so disjoint plateaus do not accumulate in bounded regions and every plateau has an immediate predecessor.

\paragraph{Construction}
For each plateau, define $S_i\coloneqq\min\{s\le a_i:L(s)=p_i\}$ (equal to $a_i$, the plateaus being maximal; the left tail and bounded closeness rule out a left-unbounded plateau, so every plateau has a finite left endpoint). Let $q_i\coloneqq\max\Bigl\{p_i-\tfrac{1}{b_i-S_i},\ p_{i-1},\ p_i-1\Bigr\}.$ Then $S_i^+\coloneqq\max\bigl\{a_i-1,\ \max\{s\in[b_{i-1},a_i]:L(s)=q_i\}\bigr\}.$ The plateaus are indexed by consecutive integers in any fixed order (legitimate by the local finiteness noted above); the modification at each plateau involves only the previous one, so no first plateau is needed (if a leftmost plateau exists, the term $p_{i-1}$ is omitted there, equivalently $b_0=-\infty$). Since $L$ is continuous and nondecreasing, with $L(a_i)=p_i$ and $L(b_{i-1})=p_{i-1}$, the target value satisfies $p_{i-1}\le q_i<p_i$ (the plateau values are strictly increasing, the plateaus being maximal intervals), so the set is nonempty and $S_i^+$ is well defined; moreover, $L(S_i^+)\ge q_i$ in either case of the maximum. On $[S_i^+,b_i]$, use the straight line segment $\ell_i(s)=\frac{p_i-L(S_i^+)}{b_i-S_i^+}(s-b_i)+p_i$ to connect $(S_i^+,L(S_i^+))$ and $(b_i,p_i)$. Define $L_{\mathrm{strict}}(s)\coloneqq\min\left(L(s),\ \ell_i(s)\right)$ on $[S_i^+,b_i]$, and set $L_{\mathrm{strict}}(s)=L(s)$ at all other points. Multiple plateaus are processed in order from left to right; all definitions refer to the original $L$ (the modification intervals being disjoint, this coincides with the partially modified function on the relevant regions).

\begin{lemma}\label{lem:strict}
The function $L_{\mathrm{strict}}$ constructed above satisfies:
\begin{enumerate}
\item $L_{\mathrm{strict}}\le L$;
\item $L_{\mathrm{strict}}$ is continuous, nondecreasing, and Lipschitz;
\item $L_{\mathrm{strict}}$ is strictly increasing;
\item Condition~\ref{cond:g_general} is preserved.
\end{enumerate}
\end{lemma}

\begin{proof}
(i) This follows directly from the definition of $\min$.

(ii) Both $L$ and $\ell_i$ are continuous, nondecreasing, and Lipschitz---the slope of $\ell_i$ is at most $\ell_{\mathrm{lip}}$: if $S_i^+=a_i-1$, then $p_i-L(S_i^+)=L(a_i)-L(a_i-1)\le\ell_{\mathrm{lip}}$ while $b_i-S_i^+=b_i-a_i+1\ge1$; if $S_i^+$ is the level point, then $p_i-L(S_i^+)\le\ell_{\mathrm{lip}}(a_i-S_i^+)$ while $b_i-S_i^+\ge a_i-S_i^+>0$; in both cases the bound follows---and the pointwise minimum preserves these properties.

(iii) On the plateau $[a_i,b_i]$, $L=p_i$ is constant, while $\ell_i$ strictly increases from a value below $p_i$ to $p_i$, so $L_{\mathrm{strict}}=\ell_i$ is strictly increasing. On non-plateau regions both $L$ and $\ell_i$ are strictly increasing (if $L(x)=L(y)$ for some $x<y$, then $L$ is constant on $[x,y]$, a plateau), hence so is their pointwise minimum. Multiple plateaus processed in order do not affect each other: the level point in the definition of $S_i^+$ lies in $[b_{i-1},a_i]$, so $S_i^+\ge b_{i-1}$ and the modification interval $[S_i^+,b_i]$ does not enter the previous plateau; hence the function is globally strictly increasing.

(iv) In this proof, $g$ and $G$ refer to the envelope pair $(L,U)$: $g(d,s)\coloneqq L(s+d)-U(s)$, $G(d)\coloneqq\inf_s g(d,s)$, and $g_{\mathrm{strict}}$, $G_{\mathrm{strict}}$ are their counterparts with $L_{\mathrm{strict}}$. The pair inherits the three conditions (Section~\ref{subsec:envelope_construction}), so $\inf_{d\ge0}G(d)>-\infty$ and $\liminf_{d\to\infty}G(d)\ge0$; the latter means that for every $\varepsilon>0$ there exists $W(\varepsilon)<\infty$ such that $G(d)\ge-\varepsilon$ for all $d>W(\varepsilon)$.

\emph{Uniform bound.} By the definition of $q_i$ and $L(S_i^+)\ge q_i$, the total modification depth satisfies $p_i-L(S_i^+)\le p_i-q_i\le\min\left\{\frac{1}{b_i-S_i},\ p_i-p_{i-1},\ 1\right\}\le1.$ For $s\in[S_i^+,b_i]$, $L(s)-L_{\mathrm{strict}}(s)\le p_i-\ell_i(s)\le p_i-L(S_i^+)\le1$, and the two functions coincide at all other points. Hence $L-1\le L_{\mathrm{strict}}\le L$, so $G_{\mathrm{strict}}(d)\ge G(d)-1\ge\inf_{d'\ge0}G(d')-1>-\infty.$ \emph{Limit inferior.} Fix $\varepsilon>0$ and set
\[
\bar d\coloneqq W(\varepsilon)+\max\left\{W(\varepsilon/2),\ \frac{2}{\varepsilon}\right\}+2.
\]
We claim that $G_{\mathrm{strict}}(d)\ge-\varepsilon$ for every $d>\bar d$. Suppose, to the contrary, that $g_{\mathrm{strict}}(d,s)<-\varepsilon$ for some $d>\bar d$ and some $s$, and write $u\coloneqq s+d$.

If $u$ lies outside every modification interval $[S_i^+,b_i]$, or lies in $[S_j^+,b_j]$ but with $\ell_j(u)\ge L(u)$, then $L_{\mathrm{strict}}(u)=L(u)$, so $g_{\mathrm{strict}}(d,s)=g(d,s)\ge G(d)\ge-\varepsilon$ since $d>\bar d\ge W(\varepsilon)$, a contradiction. Hence $u\in[S_j^+,b_j]$ for some $j$ with $\ell_j(u)<L(u)$, so that $g_{\mathrm{strict}}(d,s)=\ell_j(u)-U(s)$, and the violation reads
\begin{equation}\label{eq:strict_violation}
\ell_j(u)<U(s)-\varepsilon.
\end{equation}
Since $\ell_j$ is nondecreasing and $u\ge S_j^+$, $\ell_j(u)\ge\ell_j(S_j^+)=L(S_j^+)$. Two consequences follow. First, if $s<S_j^+$, then $g(S_j^+-s,\,s)=L(S_j^+)-U(s)\le\ell_j(u)-U(s)<-\varepsilon,$ so $S_j^+-s\le W(\varepsilon)$. Second, $u-S_j^+\le b_j-S_j^+\le(b_j-a_j)+1$, using $S_j^+\ge a_j-1$.

It remains to bound the plateau length $b_j-a_j$. Write $\gamma_j\coloneqq U(a_j)-p_j$, and note that $U$ is nondecreasing and that $S_j\le a_j$ gives $L(S_j^+)\ge q_j\ge p_j-1/(b_j-S_j)\ge p_j-1/(b_j-a_j)$.
\begin{itemize}
\item If $s\le a_j$, then $U(s)\le U(a_j)$, and
\[g_{\mathrm{strict}}(d,s)=\ell_j(u)-U(s)\ge L(S_j^+)-U(a_j)\ge-\frac{1}{b_j-a_j}-\gamma_j.\]
The violation \eqref{eq:strict_violation} forces $\gamma_j>\varepsilon/2$ or $b_j-a_j<2/\varepsilon$. If $\gamma_j>\varepsilon/2$, then for every $t\in(0,b_j-a_j)$,
\[g(t,a_j)=L(a_j+t)-U(a_j)=p_j-U(a_j)=-\gamma_j<-\varepsilon/2,\]
so $b_j-a_j\le W(\varepsilon/2)$.
\item If $s>a_j$, then $s$ lies on the plateau $[a_j,u)$. If $U(s)>p_j+\varepsilon/2$, then $g(b_j-s,s)=L(b_j)-U(s)=p_j-U(s)<-\varepsilon/2$, giving $d=u-s\le b_j-s\le W(\varepsilon/2)$. Otherwise $U(s)\le p_j+\varepsilon/2$, and $g_{\mathrm{strict}}(d,s)\ge L(S_j^+)-U(s)\ge-1/(b_j-a_j)-\varepsilon/2,$ so the violation forces $b_j-a_j<2/\varepsilon$.
\end{itemize}
In all cases, $d=(u-S_j^+)+(S_j^+-s)\le W(\varepsilon)+\max\{W(\varepsilon/2),\,2/\varepsilon\}+1<\bar d$, contradicting $d>\bar d$. Hence $G_{\mathrm{strict}}(d)\ge-\varepsilon$ for all $d>\bar d$, i.e., $\liminf_{d\to\infty}G_{\mathrm{strict}}(d)\ge0$. Condition~\ref{cond:g_general} is preserved.
\end{proof}

We close this appendix with the continuous-time counterexample, stated formally.

\begin{proposition}[No strictly increasing envelope in continuous time]\label{prop:no_strict}
There exists a pair $(\chi_1,\chi_2)$ satisfying Conditions~\ref{cond:bounded_close}, \ref{cond:large_lip}, and \ref{cond:g_continuous} that admits no strictly increasing lower envelope $L\le\chi_1$ preserving Condition~\ref{cond:g_continuous}.
\end{proposition}

\begin{proof}
On a plateau of unbounded length, any strictly increasing $L\le\chi_1$ dips strictly below the plateau value, producing a negative displacement infimum at arbitrarily large $d$ and violating Condition~\ref{cond:g_continuous}. The following pair makes this precise. Define
\[
a_i\coloneqq i(2i+1),\quad
b_i\coloneqq(i+1)(2i+1),\quad
e_i\coloneqq(i+1)(2i+3),
\]
and
\[
\chi_1(s)=\chi_2(s)=
\begin{cases}
i, & s\in[a_i,b_i),\\[6pt]
\dfrac{s-b_i}{2i+2}+i, & s\in[b_i,e_i),
\end{cases}
\]
extended as an odd function. The function $\chi_1$ has infinitely many plateaus, whose lengths $2i+1$ are unbounded. The three conditions are immediately verified: $\chi_2-\chi_1=0$ gives Condition~\ref{cond:bounded_close}; the ramp slopes $1/(2i+2)\le1/2$ give Condition~\ref{cond:large_lip}; and since $\chi_1$ is nondecreasing, $g(d,s)=\chi_1(s+d)-\chi_2(s)\ge0$ for every $d\ge0$, so $G\ge0$ identically and Condition~\ref{cond:g_continuous} holds with $d_{\max}=0$. Now suppose, for contradiction, that a strictly increasing envelope $L\le\chi_1$ exists, and take
\[
s_i=\frac{2i(2i+1)}{2},\qquad \delta_i=\frac{2i+1}{2}.
\]
Then $s_i+\delta_i=\frac{(2i+1)^2}{2}$ lies inside the plateau, and $\chi_1(s_i+\delta_i)=i$. Since $L$ is strictly increasing and $L\le\chi_1$, inside the plateau we have $L(s_i+\delta_i)<i$ (equality is excluded: $s_i+\delta_i<b_i$, so equality would force $L>i=\chi_1$ just to the right, contradicting $L\le\chi_1$), so
\[
g_L(\delta_i,s_i)=L(s_i+\delta_i)-\chi_2(s_i)<i-i=0.
\]
As $i\to\infty$, $\delta_i\to\infty$, but $g_L(\delta_i,s_i)<0$ always holds, so $\inf_s g_L(d,s)<0$ at arbitrarily large $d$, and Condition~\ref{cond:g_continuous} is violated. Hence a strictly increasing envelope does not exist in the continuous-time case.
\end{proof}

\section{Numerical examples}
\label{app:numerical}

This appendix presents three examples. Example~1 lifts the skeleton of Remark~\ref{rem:trench_rate} to a genuine Lyapunov function and displays the quantities of the theory numerically; Example~2 exhibits a closed-loop system whose comparison pair, displacement-difference infimum, and decay envelope are all known in closed form; Example~3 presents a discrete-time nonlinear regulator whose Lyapunov function has slow-growth envelopes, placing it outside every classical criterion while the discrete-time general criterion certifies it, with simulated trajectories confirming the predicted funnel.

\subsection{Example 1: the skeleton pair, lifted to a Lyapunov function}
\label{app:ex1}

Consider the skeleton $\chi_0$ of Remark~\ref{rem:trench_rate} (restated in \eqref{eq:skeleton}): the decreasing-segment lengths grow exponentially, the decreasing slopes decay double-exponentially, and the increasing segments have constant slope $2$ and length in $[0.7,1.3]$.

To lift the skeleton to a genuine Lyapunov function of $(r,\theta)$, define
\begin{equation}
V(r,\theta)=\exp\bigl(\chi_0(\ln r)+\delta(\ln r,\theta)\bigr),
\label{eq:num_perturb}
\end{equation}
where, writing $t=(s-a_j)/(a_{j+1}-a_j)$ for $s\in(a_j,a_{j+1}]$, $j\ge0$,
\begin{equation}
\begin{aligned}
\delta(s,\theta)=\;&(1-t)\frac{\rho_0\sin\theta}{(j+1)^2}+t\frac{\rho_0\sin\theta}{(j+2)^2}\\
&+\frac{\sigma_0}{(j+1)^2}\sin(\pi t)\sin(\theta+\pi t),
\end{aligned}
\label{eq:num_delta}
\end{equation}
with $\rho_0=0.15$, $\sigma_0=0.05$, and $\delta(s,\theta)=(s/a_0)\rho_0\sin\theta$ for $s\le a_0$, which makes $\delta$ continuous at $s=a_0$. Since $|\delta(s,\theta)|\le\bar u(s)$ with
\begin{equation}
\bar u(s)=(1-t)\tfrac{\rho_0}{(j+1)^2}+t\tfrac{\rho_0}{(j+2)^2}
+\tfrac{\sigma_0}{(j+1)^2}|\sin(\pi t)|
\label{eq:num_upside}
\end{equation}
for $s\in(a_j,a_{j+1}]$, and $\bar u(s)=|s|\rho_0/a_0$ for $s\le a_0$, the pair
\[
\chi_1(s)\coloneqq\chi_0(s)-\bar u(s),\qquad \chi_2(s)\coloneqq\chi_0(s)+\bar u(s)
\]
is a valid logarithmic comparison pair for $V$: $\chi_1(\ln r)\le\ln V(r,\theta)\le\chi_2(\ln r)$ for all $(r,\theta)$. This pair differs from the pair analyzed in Appendix~\ref{app:skeleton} only in the additional $\theta$-dependent term of amplitude $\sigma_0$; the analysis there applies with the amplitude bounds adjusted accordingly, and yields the same conclusions.

Fig.~\ref{fig:num_peff_LU} shows the effective power and the envelopes of the pair. In panel~(a), $p_{\mathrm{eff}}(s)=\chi_0(s)/s$ is negative on the first three decreasing segments (see the inset), becomes positive from the fourth cycle onward, and then tends to $0$ at a rate of $(\ln s)/s$, since $\chi_0(s)\approx2\ln(s/2)$ for large $s$. Panel~(b) shows the smooth monotone envelopes $L(s),U(s)$ obtained after monotonization, Lipschitzization, and mollification of the pair $(\chi_1,\chi_2)$: the gap $U-L$ attains its bound on the first few cycles (shaded band), tracks the decaying perturbation bound $2\bar u$ afterwards, and tends to $0$ across cycles---the flatter the plateau, the smaller $U-L$. This decay is precisely the geometric origin of the statement in Theorem~\ref{thm:integral_continuous} that ``$U-L$ tends to $0$ as $\tau$ increases.''

\begin{figure}[tb]
\centering
\begin{minipage}[t]{0.48\textwidth}
\centering
\includegraphics[width=\linewidth]{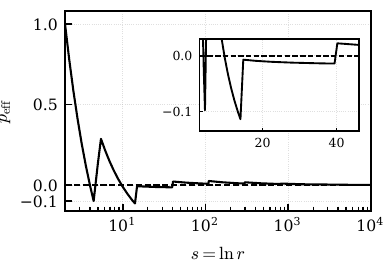}
\par\smallskip\centering\small (a)
\end{minipage}\hfill
\begin{minipage}[t]{0.48\textwidth}
\centering
\includegraphics[width=\linewidth]{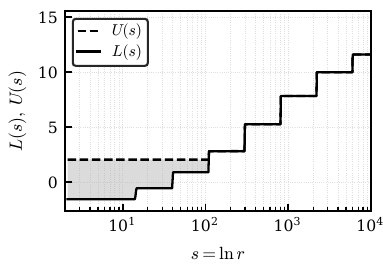}
\par\smallskip\centering\small (b)
\end{minipage}
\caption{Example 1. (a) $p_{\mathrm{eff}}(s)=\chi_0(s)/s$ (logarithmic horizontal axis); the inset magnifies the negative dips on the first three decreasing segments. $p_{\mathrm{eff}}$ changes sign and tends to $0$ at the rate $(\ln s)/s$, excluding fixed-power and same-order bounds while remaining radially unbounded. (b) The envelopes $L(s),U(s)$ after the envelope construction (logarithmic horizontal axis); the shaded band is $U-L$, which attains its bound on the first few cycles and decays to $0$ across cycles.}
\label{fig:num_peff_LU}
\end{figure}

\paragraph{Failure of classical criteria}
All conventional criteria fail for this pair. Hahn-type criteria require bounds of a fixed power or of the same order, which is excluded by the sign-changing, vanishing $p_{\mathrm{eff}}$. For the Panteley--Lor\'ia criterion, the first inequality of \eqref{eq:panteley_conditions}, i.e., $\chi_2(r)\le\chi_1(r+\ln c_1)$, fails for every constant $c_1$: for $c_1>1$, at $r=a_j$ with $j$ so large that $b_j-a_j>\ln c_1$, the point $r+\ln c_1$ still lies on the $j$-th decreasing segment, and hence
\[
\chi_1(r+\ln c_1)-\chi_2(r)=-e^{-j^2}\ln c_1-\bar u(r+\ln c_1)-\bar u(r)<0.
\]
For $0<c_1<1$, take $r=b_j$ with $|\ln c_1|<D_j$: then $r+\ln c_1\in(a_j,b_j)$ still lies on the decreasing segment, and
\[
\chi_1(r+\ln c_1)-\chi_2(r)=e^{-j^2}|\ln c_1|-\bar u(r+\ln c_1)-\bar u(r)<0
\]
for all sufficiently large $j$, since $e^{-j^2}|\ln c_1|=o(j^{-2})$ while $\bar u\asymp A/j^2$ on cycle $j$. (The case $c_1=1$ fails immediately, since $\chi_2>\chi_1$ wherever $\bar u>0$.)
(The second inequality holds with an explicit constant $\gamma$---by bounded closeness and Lipschitzness, $c_2=e^{-(D+\Lambda c)}$ and $c_3=e^{-c}$ serve for any $c>0$, as in the converse direction of Proposition~\ref{prop:pl_equiv}---so the Panteley--Lor\'ia criterion fails only through its first inequality; moreover, by Proposition~\ref{prop:no_strict}, the strictly increasing envelopes required for its application need not exist in the continuous-time setting.) On the other hand, Conditions~\ref{cond:bounded_close} and \ref{cond:large_lip} hold: $\chi_2-\chi_1=2\bar u\le2(\rho_0+\sigma_0)$ is uniformly bounded, and both $\chi_0$ and $\bar u$ have uniformly bounded slopes, so $\chi_1$ is globally Lipschitz.

\paragraph{The integral tool}
Fig.~\ref{fig:num_tau_gap} illustrates the integral tool on this pair (logarithmic horizontal axes). Panel~(a) plots $\tau(s)$: it takes larger values on the decreasing segments (plateau positions) and smaller values on the increasing segments, reflecting the local shape of $L$. Panel~(b) plots $\min_s 1/(U(s)-L(s))$ against $\tau$: the curve increases monotonically, indicating that $U-L\to0$ as $\tau\to\infty$, consistently with the necessary condition of Theorem~\ref{thm:integral_continuous}.

\begin{figure}[tb]
\centering
\begin{minipage}[t]{0.48\textwidth}
\centering
\includegraphics[width=\linewidth]{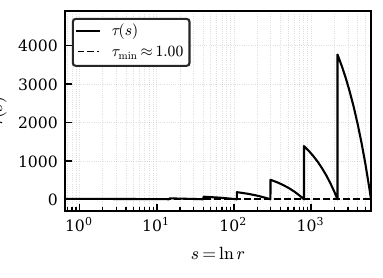}
\par\smallskip\centering\small (a)
\end{minipage}\hfill
\begin{minipage}[t]{0.48\textwidth}
\centering
\includegraphics[width=\linewidth]{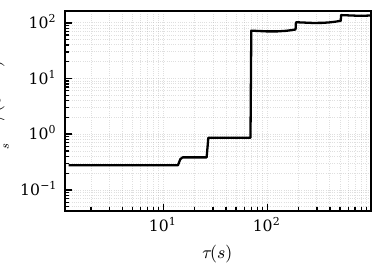}
\par\smallskip\centering\small (b)
\end{minipage}
\caption{The integral tool for Example 1 (logarithmic horizontal axes). (a) $\tau(s)$: larger values at plateau positions, smaller values on increasing segments. (b) $\min_s 1/(U(s)-L(s))$ versus $\tau$ (log-log; $0.1\%$--$99.9\%$ quantile-truncated, block-minimum downsampled): monotone increase, i.e., $U-L\to0$ as $\tau\to\infty$.}
\label{fig:num_tau_gap}
\end{figure}

\paragraph{Closed-form analysis of $G(d)$}
Let $L_0,U_0$ be the envelopes of the pair $(\chi_1,\chi_2)$ produced by the monotonization and Lipschitzization steps of Section~\ref{subsec:envelope_construction} (continuous, nondecreasing, $4$-Lipschitz, so the pair conventions hold without mollification; the mollification with width $\sigma_m=0.01$ shifts each envelope by at most $4\sigma_m$ uniformly, and is immaterial here). Define $G(d)=\inf_s[L_0(s+d)-U_0(s)]$. By Proposition~\ref{prop:skeleton}, $G(d)<0$ for every $d>0$, $G_{\min}$ is finite (numerically $G_{\min}\approx-3.77$, attained on the first decreasing segment), and $-G(d)=\Theta\bigl((\ln d)^{-2}\bigr)$ as $d\to\infty$, so $G_{\lim}=0$. Hence Conditions~\ref{cond:bounded_close}, \ref{cond:large_lip}, and \ref{cond:g_general} hold while Condition~\ref{cond:g_continuous} fails: by Theorems~\ref{thm:general} and \ref{thm:continuous}, the discrete-time general criterion covers this pair and the continuous-time general criterion rejects it, so the skeleton realizes the separation between the discrete-time and continuous-time criteria. Fig.~\ref{fig:num_G} displays the computed $G(d)$: panel~(a) shows the negative plateaus $\approx-3.77,-2.72,-1.27$, separated by the transition points $b_j-a_0$; panel~(b) confirms the $(\ln d)^{-2}$ decay of $-G(d)$ to $0^{-}$; numerically, $0.28\le-G(d)(\ln d)^2\le0.33$ for $d\in[10^3,10^8]$.

\begin{figure}[tb]
\centering
\begin{minipage}[t]{0.48\textwidth}
\centering
\includegraphics[width=\linewidth]{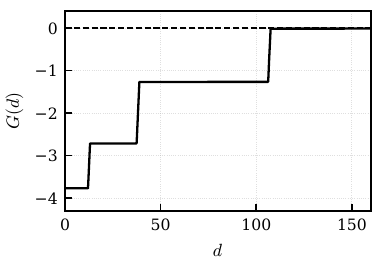}
\par\smallskip\centering\small (a)
\end{minipage}\hfill
\begin{minipage}[t]{0.48\textwidth}
\centering
\includegraphics[width=\linewidth]{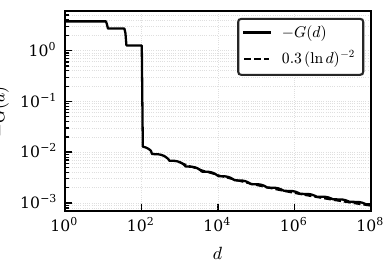}
\par\smallskip\centering\small (b)
\end{minipage}
\caption{The infimum $G(d)=\inf_s[L_0(s+d)-U_0(s)]$ for the skeleton pair. (a) Small and moderate $d$: negative piecewise-constant plateaus separated by $b_j-a_0$. (b) Large $d$ (log-log coordinates): $-G(d)$ decays at the rate $(\ln d)^{-2}$ (dashed reference). Hence $G(d)<0$ for every $d$ yet $G_{\lim}=0$: the discrete-time conditions hold and the continuous-time one fails.}
\label{fig:num_G}
\end{figure}

\paragraph{Summary}
The skeleton thus provides a second, mechanism-wise independent realization of the separation of Theorem~\ref{thm:continuous}: for the fast-decay pair of Section~\ref{subsec:ex_fastdecay} it is driven by the diminishing increments of a slow-growth lower function, whereas here by double-exponentially flat decreasing segments with a directional perturbation keeping $G(d)$ negative for every $d$ while letting it tend to $0$.

\subsection{Example 2: a realizable closed-loop system}
\label{app:ex2}

The preceding example works with a synthesized pair. We now exhibit a genuine closed-loop system whose comparison pair, displacement-difference infimum, and decay envelope are all known in closed form.

Take $\alpha_1(r)=r^2$ and $\alpha_2(r)=c(r)\,r^2$, where $c\in C^2(\Rnn)$ satisfies $c(r)=2$ for $r\le1$, $c(r)=3$ for $r\ge2$, and $c'(r)\ge0$ on $[1,2]$ (e.g., $c(r)=2+p(r-1)$ on $[1,2]$ with $p(t)=6t^5-15t^4+10t^3$). Define, for $x=(r\cos\theta,r\sin\theta)$,
\[
V(x)=\alpha_1(r)+\bigl(\alpha_2(r)-\alpha_1(r)\bigr)\cos^2\theta
=r^2+\bigl(c(r)-1\bigr)x_1^2,
\]
\[
f(x)=-\beta V(x)\,\frac{\nabla V(x)}{|\nabla V(x)|^2}\quad(x\neq0),\qquad f(0)=0,
\]
with $\beta=1$. By Remark~\ref{rem:realizability}, the radial bounds of $V$ are exactly $\alpha_1,\alpha_2$. Near the origin $V=r^2+x_1^2$ is quadratic, so $|\nabla V(x)|\ge\kappa\norm{x}$ for some $\kappa>0$ and $|f(x)|\le(\beta/\kappa)V(x)/\norm{x}\to0$ as $x\to0$: the field is continuous on $\R^2$ and smooth off the origin. Along every solution, $\dot V=-V$; solutions are unique (the field is locally Lipschitz off the origin, and a solution reaching the origin satisfies $V\equiv0$) and global (the sublevel sets of $V$ are forward invariant and bounded). The logarithmic comparison pair is
\[
\chi_1(s)=2s,\qquad \chi_2(s)=2s+\ln c(e^s).
\]
Then $\chi_2-\chi_1=\ln c(e^s)\in[\ln2,\ln3]$ (Condition~\ref{cond:bounded_close}); $\chi_2'(s)=2+(c'/c)(e^s)\,e^s$ is uniformly bounded, since $c'$ is supported on $r\in[1,2]$ (Condition~\ref{cond:large_lip}); and
\[
G(d)=\inf_{s}\bigl[\chi_1(s+d)-\chi_2(s)\bigr]=2d-\ln3,
\]
so $G(d)\ge0$ for all $d\ge d_{\max}\coloneqq\tfrac12\ln3$, and Condition~\ref{cond:g_continuous} holds. By Theorem~\ref{thm:continuous} the origin is exponentially stable, and the criterion is decisive in both time settings. (This pair is also covered by the classical criteria---the bounds have the same power $2$, and \eqref{eq:panteley_conditions} holds with $c_1=\sqrt3$---which is precisely the point: on the realizable subclass the conditions do not collapse, and they recover the sharp transient constants below.)

The example also displays the quantitative content of the criterion. Since $V(x(t))=e^{-t}V(x_0)$ and $r^2\le V\le3r^2$,
\begin{equation}
\sqrt{V(x_0)/3}\;e^{-t/2}\le\norm{x(t)}\le\sqrt{V(x_0)}\;e^{-t/2},
\label{eq:num_funnel}
\end{equation}
i.e., the decay rate is $\beta/2$ and the transient overshoot factor is $\sqrt3=e^{\sup(\chi_2-\chi_1)/2}$, exactly the constants tracked by the explicit rate of Remark~\ref{rem:rate}. Fig.~\ref{fig:num_traj}(a) shows the simulated trajectories together with the level sets of $V$; the trajectories bend toward the $x_2$-axis, along which $V$ is minimal for a given radius. Fig.~\ref{fig:num_traj}(b) plots the normalized radius $\norm{x(t)}e^{t/2}/\sqrt{V(x_0)}$: all four trajectories remain inside the funnel $[1/\sqrt3,1]$ predicted by \eqref{eq:num_funnel}, and the trajectory starting on the $x_1$-axis attains the lower boundary $1/\sqrt3$ at $t=0$, showing that the constant $\sqrt3$ is sharp. The subsequent rise of that curve from $1/\sqrt3$ toward $1/\sqrt2$ reflects the tightening of the ratio $c(r)$ from $3$ to $2$ as $r(t)$ crosses below $1$.

\begin{figure}[tb]
\centering
\begin{minipage}[t]{0.48\textwidth}
\centering
\includegraphics[width=\linewidth]{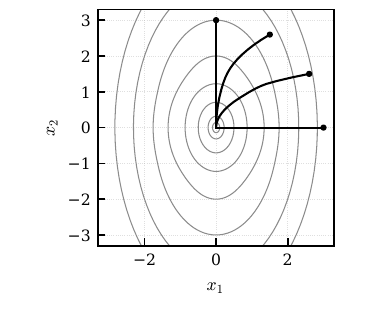}
\par\smallskip\centering\small (a)
\end{minipage}\hfill
\begin{minipage}[t]{0.48\textwidth}
\centering
\includegraphics[width=\linewidth]{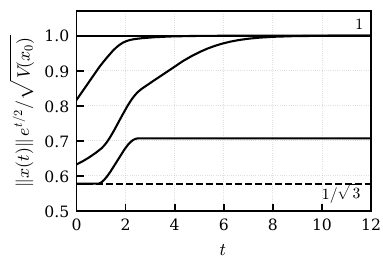}
\par\smallskip\centering\small (b)
\end{minipage}
\caption{Example 2: the realizable closed-loop system. (a) Trajectories from four initial conditions with $\norm{x_0}=3$ and $\theta_0\in\{0,\pi/6,\pi/3,\pi/2\}$ (dots), with the level sets of $V$ (gray). (b) Normalized radius $\norm{x(t)}e^{t/2}/\sqrt{V(x_0)}$: all trajectories stay inside the predicted funnel $[1/\sqrt3,\,1]$ (dashed), confirming the decay rate $\beta/2$ and the sharp overshoot factor $\sqrt3$.}
\label{fig:num_traj}
\end{figure}

\subsection{Example 3: a discrete-time nonlinear regulator with a slow-growth Lyapunov function}
\label{app:ex3}

We close with a genuine discrete-time closed-loop system whose comparison pair consists of slow-growth logarithmic functions: it lies outside every classical criterion, yet the discrete-time general criterion certifies it, with all data in closed form.

\paragraph{System and Lyapunov function}
Consider the planar discrete-time system
\begin{equation}
x_{k+1}=\frac{\psi(\norm{x_k})}{\norm{x_k}}\,x_k,\qquad \psi(r)\coloneqq\sqrt{\exp\Bigl(\sqrt{\ln(e+r^2)}\Bigr)-e},
\label{eq:num_slowsys}
\end{equation}
arising, e.g., from a nonlinear regulator whose gain is scheduled by the amplitude. The map preserves directions and contracts the radius by the state-dependent gain $\psi(r)/r$: the gain stays below $0.9$ everywhere, $\psi(r)/r\to1/\sqrt2$ as $r\to0$, and $\psi(r)=r^{o(1)}$ as $r\to+\infty$, so the contraction is geometric near the origin and super-geometric for large states. Take the Lyapunov function
\begin{equation}
V(x)=\ln\ln(e+\norm{x}^2)+(\ln2)\,\frac{\norm{x}^2}{1+\norm{x}^2}\cos^2\theta,
\label{eq:num_slowV}
\end{equation}
where $\theta$ is the polar angle of $x$. Writing $\ell(r)\coloneqq\ln\ln(e+r^2)$ and $h(r)\coloneqq r^2/(1+r^2)$, one step of \eqref{eq:num_slowsys} gives
\[
V(x_{k+1})=\tfrac12\,\ell(\norm{x_k})+(\ln2)\,h(\psi(\norm{x_k}))\cos^2\theta_k,
\]
which is at most $0.7\,V(x_k)$: the constant $0.7$ is certified by a one-dimensional maximization, as the ratio $V(x_{k+1})/V(x_k)$ is continuous in $r$, tends to $1/2$ both as $r\to0$ and as $r\to+\infty$, and a mesh evaluation with a Lipschitz certificate gives the supremum $0.69$. Hence the decay premise holds with $\beta=-\ln 0.7\approx0.36$.

\paragraph{Failure of the classical criteria}
The radial envelopes are
\[
\alpha_1(X)=\ln\ln(e+X^2),\qquad \alpha_2(X)=\alpha_1(X)+(\ln2)\frac{X^2}{1+X^2},
\]
so $\chi_i(s)=\ln\alpha_i(e^s)\sim\ln\ln(2s)$ as $s\to+\infty$: a slow-growth pair. Since $\alpha_1(X)/X^p\to0$ as $X\to+\infty$ for every $p>0$, no power lower bound exists, and Hahn's same-order criterion does not apply. For the Panteley--Lor\'ia criterion, $\alpha_1^{-1}(y)=\sqrt{\exp(e^y)-e}$, and since $\alpha_2(s)=\alpha_1(s)+\ln2+o(1)$ as $s\to+\infty$,
\[
\alpha_1^{-1}(\alpha_2(s))=\sqrt{(e+s^2)^{2}-e}\sim s^{2},
\]
which is superlinear: the first inequality in \eqref{eq:panteley_conditions} fails.

\paragraph{Verification of the conditions}
Since $\chi_2(s)-\chi_1(s)=\ln\bigl(1+(\ln2)h(e^s)/\alpha_1(e^s)\bigr)$, a one-dimensional maximization gives $\chi_2-\chi_1\le D\approx1.07$ (Condition~\ref{cond:bounded_close}). The slope of $\chi_1$ satisfies $\chi_1'(s)\le2.02$ for all $s$, so Condition~\ref{cond:large_lip} holds. For the joint lower bound, $G(d)\ge-D$ since $\chi_1$ is nondecreasing. As for the fast-decay pair of Section~\ref{subsec:ex_fastdecay}, the gap term dominates the increment in the far tail: for every fixed $d>0$, taking $s$ large gives $g(d,s)<0$, so $G(d)<0$ for all $d>0$ and the strengthened Condition~\ref{cond:g_continuous} fails. On the other hand, balancing the two rates gives $G(d)\to0^{-}$ as $d\to+\infty$, so $G_{\min}=G(0)=-D>-\infty$ and $G_{\lim}=0$: Conditions~\ref{cond:package} and \ref{cond:g_general} hold. By Theorem~\ref{thm:general}, the discrete-time general criterion certifies exponential stability, while the pair falls in the separation regime of Theorem~\ref{thm:continuous}: no uniform continuous-time guarantee exists for this pair.

\paragraph{Simulation}
Fig.~\ref{fig:num_slow} shows the iteration of \eqref{eq:num_slowsys} from four initial conditions: panel~(a) displays the iterates together with the level sets of $V$; panel~(b) plots $\norm{x_k}$ against $k$ (log scale) together with the certified funnel $\norm{x_k}\le\alpha_1^{-1}\bigl(0.7^k\alpha_2(\norm{x_0})\bigr)$ furnished by the decay premise and the envelopes (the quantitative content tracked by Remark~\ref{rem:rate}). The decay is geometric with rate about $1/\sqrt2$ per step in the mid range and accelerates for large states, matching the $V$-space contraction.

\begin{figure}[tb]
\centering
\begin{minipage}[t]{0.48\textwidth}
\centering
\includegraphics[width=\linewidth]{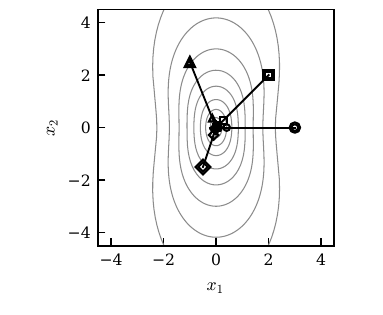}
\par\smallskip\centering\small (a)
\end{minipage}\hfill
\begin{minipage}[t]{0.48\textwidth}
\centering
\includegraphics[width=\linewidth]{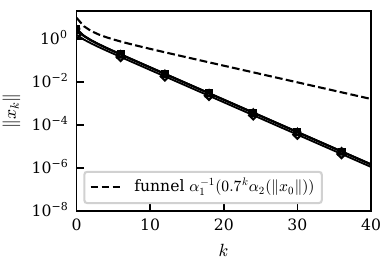}
\par\smallskip\centering\small (b)
\end{minipage}
\caption{Example 3: the discrete-time nonlinear regulator with a slow-growth Lyapunov function. (a) Iterates from four initial conditions, with the level sets of $V$ (gray). (b) Norm decay: $\norm{x_k}$ (log scale) for the four initial conditions, with the certified funnel $\alpha_1^{-1}(0.7^k\alpha_2(\norm{x_0}))$ (dashed).}
\label{fig:num_slow}
\end{figure}

\subsection{Computational notes}
\label{app:compnotes}
All objects in this appendix are closed-form, and every figure and numerical statement is reproducible by direct evaluation. The skeleton quantities of Example~1 were computed on a uniform $s$-mesh of $40000$ points over $[\ln(0.01),\,10^4]$, with the envelope stages of Section~\ref{subsec:envelope_construction} implemented exactly as defined (mollification width $\sigma_m=0.01$); the values of $G(d)$ were obtained by direct minimization over the mesh, and the bracket $0.28\le-G(d)(\ln d)^2\le0.33$ was read off a log-spaced mesh over $[10^3,10^8]$. The decay constant of Example~3 is certified by evaluating the ratio $V(x_{k+1})/V(x_k)$---continuous in $r$, with limit $1/2$ at both ends---on a mesh of a bounded interval covering the transition region, the mesh supremum plus a Lipschitz remainder staying below $0.7$. The trajectories of Example~2 were integrated with a standard adaptive Runge--Kutta solver.

\section{The strict decay premise and the level set}
\label{app:beta}

The level-set formulation of Section~\ref{subsec:analysis} rests on the strict inequality: the level-set supremum $\Gamma_\beta(s_0,t)=\sup\{s:\chi_1(s)<\chi_2(s_0)-\beta t\}$ does not cover states at which equality $V(x)=e^{-\beta t}V(x_0)$ holds exactly. Accordingly, the strict $\beta$-criterion is the constituent analyzed in Sections~\ref{sec:conditions}--\ref{sec:continuous}: for the non-strict $\beta$-criterion the equality case is feasible, and the transfer from the strict constituent to the general criterion is effected by the inclusion argument of Section~\ref{subsec:analysis} ($\Xi_{2\beta}\subseteq\Xi_\beta^{\mathrm{s}}$ away from the trivial trajectory), not by the level set directly.

\fi

\paragraph*{Declarations} During the preparation of this manuscript, the authors used a large language model (Kimi, Moonshot AI) for English polishing, editorial suggestions, proof checking, and figure assistance; it was not used to generate the core scientific ideas, main results, or proof strategies. The authors assume responsibility for all content.

\end{document}